\documentclass[]{interact}

\usepackage{epstopdf}% To incorporate .eps illustrations using PDFLaTeX, etc.
\usepackage[caption=false]{subfig}% Support for small, `sub' figures and tables

\usepackage[longnamesfirst,sort]{natbib}% Citation support using natbib.sty
\bibpunct[, ]{(}{)}{;}{a}{,}{,}% Citation support using natbib.sty
\renewcommand\bibfont{\fontsize{10}{12}\selectfont}% To set the list of references in 10 point font using natbib.sty

\theoremstyle{plain}% Theorem-like structures provided by amsthm.sty
\newtheorem{theorem}{Theorem}[section]

\theoremstyle{definition}

\theoremstyle{remark}
\newtheorem{remark}{Remark}

\usepackage[dvipsnames]{xcolor}

\usepackage{nomencl}
\usepackage{ifthen}
\renewcommand{\nomgroup}[1]{%
	\ifthenelse{\equal{#1}{C}}{\item[\textbf{Calligraphic Symbols}]}{%
		\ifthenelse{\equal{#1}{G}}{\item[\textbf{Greek Symbols}]}{%
			\ifthenelse{\equal{#1}{R}}{\item[\textbf{Roman Symbols}]}{}}}
}
\makenomenclature

\begin{document}

%\articletype{ARTICLE TEMPLATE}% Specify the article type or omit as appropriate

\title{Modeling, Control and Human-In-The-Loop Stability Analysis of an Elastic Quadrotor}

\author{
\name{Emre Eraslan\textsuperscript{a}\thanks{CONTACT Emre Eraslan. Email: emre.eraslan@bilkent.edu.tr} and Yildiray Yildiz\textsuperscript{a}}
\affil{\textsuperscript{a}Mechanical Engineering, Bilkent University, Cankaya, Ankara 06800, Turkey (e-mail: \{emre.eraslan, yyildiz\}@bilkent.edu.tr).}
}

\maketitle

\begin{abstract}
 This paper introduces an analytical framework for the derivation of hybrid equations of motion of a flexible quadrotor. This approach helps
 obtain rigid and elastic equations of motion simultaneously, in a decoupled form, which facilitates the controller design. A delay-dependent stability condition is obtained for the overall system dynamics, including the operator with reaction time delay, the adaptive controller and the flexible quadrotor dynamics. Two different adaptive controllers are implemented to address the uncertainties. It is demonstrated via simulations that the flexible arm tip oscillations are mitigated when a closed loop reference model adaptive controller is used, compared to a conventional model reference adaptive controller.   
 \end{abstract}

\nomenclature[C]{$\mathcal{A}_{d}$}{coefficient matrix of delayed term of the delay equation}
\nomenclature[C]{$\mathcal{A}_{n}$}{coefficient matrix of the undelayed term of the delay equation}
\nomenclature[C]{$\mathcal{D}$}{Rayleigh dissipation function}
\nomenclature[C]{$\mathcal{F}$}{inertial frame reference}
\nomenclature[C]{$\mathcal{G}$}{non-inertial frame reference}
\nomenclature[C]{$\mathcal{L}$}{Lagrangian}
\nomenclature[C]{$\mathcal{M}{_e}$}{tracking error performance metric}
\nomenclature[C]{$\mathcal{Q}_i$}{generalized force for the $i^{\mathrm{th}}$ generalized coordinate}
\nomenclature[C]{$\mathcal{Q}_\xi$}{overall thrust vector in the inertial frame}
\nomenclature[C]{$\mathcal{Q}_\upsilon$}{overall torque vector in the inertial frame}
\nomenclature[C]{$\mathcal{Q}_{\Upsilon_j}$}{$j^{\mathrm{th}}$ generalized force for the generalized displacement coordinate}
\nomenclature[C]{$\mathcal{R_E}$}{inertial position of a mass element}
\nomenclature[C]{$\mathcal{R_F}$}{position of body reference frame relative to inertial frame}
\nomenclature[C]{$\mathcal{R_G}$}{position relative to a non-inertial reference frame}
\nomenclature[C]{$\mathcal{T}$}{kinetic energy}
\nomenclature[C]{$\mathcal{T}_e$}{elastic kinetic energy}
\nomenclature[C]{$\mathcal{T}_r$}{rotational kinetic energy}
\nomenclature[C]{$\mathcal{T}_t$}{translational kinetic energy}
\nomenclature[C]{$\mathcal{U}$}{potential energy}
\nomenclature[C]{$\mathcal{U}_e$}{elastic potential energy}
\nomenclature[C]{$\mathcal{U}_g$}{gravitational potential energy}
\nomenclature[C]{$\mathcal{V}$}{Lyapunov function candidate}
\nomenclature[G]{$\beta_j$}{constant obtained from the partial differential equation for the $j^{\mathrm{th}}$ mode}
\nomenclature[G]{$\bar{\beta}_j$}{solution of the transcendental frequency equation for the $j^{\mathrm{th}}$ mode}
\nomenclature[G]{$\bar{\beta}_j^*$}{ratio of hyperbolic functions of $\bar{\beta}_j$ for the $j^{\mathrm{th}}$ mode}
\nomenclature[G]{$\gamma_c$}{constant related to the normalization constant of $W_j$}
\nomenclature[G]{$\bar{\gamma}_j$}{normalization constant corresponding to the $j^{\mathrm{th}}$ mode}
\nomenclature[G]{$\Gamma_{\Theta_{ii}}$}{adaptation rate  for the $i^{\mathrm{th}}$ diagonal}
\nomenclature[G]{$\delta$}{Dirac's delta function}
\nomenclature[G]{$\delta_{jl}$}{Kronecker delta}
\nomenclature[G]{$\zeta$}{input to the human dynamics}
\nomenclature[G]{$\eta$}{human state vector}
\nomenclature[G]{$\Theta$}{unknown overall weight matrix}
\nomenclature[G]{$\Theta_p$}{unknown weight matrix}
\nomenclature[G]{$\hat{\Theta}$}{time-varying adaptative parameter matrix}
\nomenclature[G]{$\tilde{\Theta}$}{adaptive parameter estimation error}
\nomenclature[G]{$\Lambda$}{control effectiveness matrix}
\nomenclature[G]{$\mu$}{regressor vector regarding the delay differential equation}
\nomenclature[G]{$\xi$}{position variable vector}
\nomenclature[G]{$\Pi$}{nonhomogeneous part of the delay differential equation}
\nomenclature[G]{$\rho$}{density of mass element}
\nomenclature[G]{$\rho_c$}{density of quadrotor arm}
\nomenclature[G]{$\sigma$}{damping coefficient term}
\nomenclature[G]{$\sigma_c$}{damping coefficient of quadrotor arm}
\nomenclature[G]{$\sigma_c'$}{normalized damping coefficient of quadrotor arm}
\nomenclature[G]{$\tau_b$}{torque vector in the inertial frame}
\nomenclature[G]{$\tau_g$}{gyroscopic torque}
\nomenclature[G]{$\tau_h$}{reaction delay}
\nomenclature[G]{$\tau_m$}{the smallest time constant of the reference model}
\nomenclature[G]{$\tau_\theta$}{pitch torque}
\nomenclature[G]{$\tau_\phi$}{roll torque}
\nomenclature[G]{$\tau_\psi$}{yaw torque}
\nomenclature[G]{$\upsilon$}{Euler angles}
\nomenclature[G]{$\Upsilon_j$}{generalized displacement coordinates for the $j^{\mathrm{th}}$ mode}
\nomenclature[G]{$\Upsilon_{kj}$}{generalized displacement coordinates on the $k^{\mathrm{th}}$ arm for the $j^{\mathrm{th}}$ mode}
\nomenclature[G]{$\Phi$}{high order nonlinear effect matrix}
\nomenclature[G]{$\omega$}{angular velocity vector}
\nomenclature[G]{$\bar{\omega}_j$}{natural frequency for the $j^{\mathrm{th}}$ mode}
\nomenclature[G]{$\bar{\omega}_{kj}$}{natural frequency on the $k^{\mathrm{th}}$ arm for the $j^{\mathrm{th}}$ mode}
\nomenclature[G]{$\Omega_k$}{angular velocity of the $k^{\mathrm{th}}$ arm}
\nomenclature[G]{$\Omega_g$}{gyroscopic velocity}
\nomenclature[G]{$\Omega_s$}{angular velocity squared vector}
\nomenclature[R]{$A$}{extended open loop system matrix}
\nomenclature[R]{$A_c$}{cross-sectional area of quadrotor arm}
\nomenclature[R]{$A_e$}{elastic mode system matrix}
\nomenclature[R]{$A_h$}{human operator system matrix}
\nomenclature[R]{$A_m$}{reference model system matrix}
\nomenclature[R]{$A_p$}{system matrix}
\nomenclature[R]{$A'_e$}{elastic mode system matrix for the $k^{\mathrm{th}}$ arm}
\nomenclature[R]{$B$}{extended open loop input matrix}
\nomenclature[R]{$B_{e}$}{elastic mode input matrix for $u$}
\nomenclature[R]{$B_h$}{human operator input matrix}
\nomenclature[R]{$B_m$}{reference model input matrix}
\nomenclature[R]{$B_p$}{input matrix}
\nomenclature[R]{$B_{ze}$}{elastic mode input matrix for $F$}
\nomenclature[R]{$B'_{ze}$}{elastic mode input matrix for $F$ for the $k^{\mathrm{th}}$ arm}
\nomenclature[R]{$c$}{command to the human operator}
\nomenclature[R]{$C_h$}{human operator output matrix}
\nomenclature[R]{$C_p$}{output matrix}
\nomenclature[R]{$D_h$}{feedforward human operator matrix}
\nomenclature[R]{$e$}{reference model tracking error}
\nomenclature[R]{$e_p$}{integrated tracking error}
\nomenclature[R]{$E_c$}{Young modulus of quadrotor arm}
\nomenclature[R]{$E_h$}{state subset selector matrix}
\nomenclature[R]{$F$}{force vector in the inertial frame}
\nomenclature[R]{$F_b$}{total thrust force in the body frame}
\nomenclature[R]{$F_k$}{thrust force on the $k^{\mathrm{th}}$ arm}
\nomenclature[R]{$g$}{gravitational acceleration}
\nomenclature[R]{$G_{\mathrm{PI}}$}{transfer function of human operator}
\nomenclature[R]{$h_j$}{$j^{\mathrm{th}}$ convex function}
\nomenclature[R]{$H$}{convex function vector}
\nomenclature[R]{$i$}{generalized coordinate counter}
\nomenclature[R]{$I$}{moment of inertia matrix}
\nomenclature[R]{$j$}{mode counter}
\nomenclature[R]{$J_c$}{moment of inertial of quadrotor arm}
\nomenclature[R]{$J_r$}{moment of inertia rotor}
\nomenclature[R]{$J_x$}{moment of inertia in the $x$ direction}
\nomenclature[R]{$J_y$}{moment of inertia in the $y$ direction}
\nomenclature[R]{$J_z$}{moment of inertia in the $z$ direction}
\nomenclature[R]{$k$}{quadrotor arm counter}
\nomenclature[R]{$k_q$}{drag factor}
\nomenclature[R]{$k_t$}{thrust factor}
\nomenclature[R]{$K$}{state feedback control gain matrix}
\nomenclature[R]{$K_p$}{integral gain of human operator transfer function}
\nomenclature[R]{$L$}{CRM error coefficient matrix}
\nomenclature[R]{$L_c$}{length of quadrotor arm}
\nomenclature[R]{$m$}{total quadrotor mass}
\nomenclature[R]{$m_b$}{main body mass}
\nomenclature[R]{$m_c$}{mass of quadrotor arm}
\nomenclature[R]{$m_r$}{rotor mass}
\nomenclature[R]{$\bar{m}$}{ratio of rotor mass to mass of quadrotor arm}
\nomenclature[R]{$M_j$}{modal mass corresponding to the $j^{\mathrm{th}}$ mode}
\nomenclature[R]{$p$}{truncated elastic degrees of freedom}
\nomenclature[R]{$P$}{solution of the Lyapunov equation}
\nomenclature[R]{$q$}{generalized coordinate}
\nomenclature[R]{$Q$}{free parameter in the Lyapunov equation}
\nomenclature[R]{$r$}{reference formed by the human operator}
\nomenclature[R]{$\bar{r}$}{maximum reference value}
\nomenclature[R]{$R^{B}$}{rotation matrix from $\mathcal{G}$ to $\mathcal{F}$}
\nomenclature[R]{$R^{F}$}{rotation matrix from $F$ to $u$}
\nomenclature[R]{$R^{\upsilon}$}{rotation matrix from $\upsilon$ to $\omega$}
\nomenclature[R]{$R^{\Omega_s}$}{rotation matrix from $\Omega_s$ to $u$}
\nomenclature[R]{$s$}{Laplace operator}
\nomenclature[R]{$\bar{s}$}{constant undeformed length}
\nomenclature[R]{$t$}{time}
\nomenclature[R]{$t_a$}{anomaly time}
\nomenclature[R]{$T$}{simulation time}
\nomenclature[R]{$T_p$}{time constant of human operator transfer function}
\nomenclature[R]{$u$}{control input}
\nomenclature[R]{$u_{\mathrm{ad}}$}{adaptive control input}
\nomenclature[R]{$u_{\mathrm{bl}}$}{baseline control input}
\nomenclature[R]{$w$}{relative displacement}
\nomenclature[R]{$w_k$}{relative displacement for the $k^{\mathrm{th}}$ arm}
\nomenclature[R]{$W_j$}{mode shape for the $j^{\mathrm{th}}$ mode}
\nomenclature[R]{$x$}{extended state vector}
\nomenclature[R]{$x_m$}{reference model vector}
\nomenclature[R]{$x_p$}{system state vector}
\nomenclature[R]{$\bar{x}$}{coordinate on quadrotor arm}
\nomenclature[R]{$z_e$}{elastic state vector}
\nomenclature[R]{$z_e^k$}{elastic state vector on the $k^{\mathrm{th}}$ arm}
\nomenclature[R]{$z_{kj}$}{elastic state on the $k^{\mathrm{th}}$ arm for the $j^{\mathrm{th}}$ mode}
\printnomenclature


\section{Introduction}
Aerial vehicles are conventionally formulated as rigid bodies. However, modeling of elastic effects can contribute significantly to the dynamic response of the vehicle. Despite the usual practice of separating the dynamic analysis of aircraft into rigid and elastic dynamics \citep{rasti12}, a large body of literature has been devoted to hybrid equations of motion \citep{mei66,wasz88,flatus92}. In particular, a number of approaches to flexible aircraft design have been proposed, such as aerodynamic strip theory on wings \citep{wasznasa87}, bifurcation and continuation methods \citep{bagh2011}, nonlinear reduced order models \citep{daronch2012}, interconnected multiple beam structure method \citep{chang2008}, and structural dynamic modeling method  
\citep{ngu2009}. These studies are also extended to UAVs that have long flexible arms \citep{vanschoor90,ritt2016,cesnik2010}. On the other hand, literature on elastic dynamics pertaining to quadrotor UAVs is relatively scarce. In \citet{srikant2010}, quadrotor flexibility is formulated as shape-shifting of the quadrotor chassis upon impact with a wall. In some other approaches, structural vibration analysis of a quadrotor is conducted and experimentally verified \citep{verbeke2016,tullu2018}. These methods generally focus only on the flexible effects and do not shed light on rigid and elastic dynamics as a whole.

In this paper, we introduce an analytical framework to derive hybrid equations of motion of a flexible quadrotor. The applied method is a comprehensive procedure predicated on Lagrangian mechanics using the mean-axes theorem. This approach helps obtain rigid and elastic equations of motion simultaneously, in a decoupled form, which facilitates the controller design. To compensate for the uncertainty sources such as flight anomalies, actuator failures and model linearization effects, we implement two different adaptive controllers to control the flexible UAV: One of them is the conventional model reference adaptive controller (MRAC) \citep{narendra2012}, and the other is the closed loop reference model (CRM) adaptive controller \citep{step2010a,step2011b,lav2011,gibson2012,gibson2013a,gibson2013b,gibsonphd2014,yucelen2014}. CRM adaptive controller is developed to reduce the oscillations in MRAC architectures, and its effectiveness is verified experimentally \citep{alan2018,eraslan2019}. We show that CRM adaptive controller indeed helps reduce the vibrations of the flexible quadrotor arms. Finally, we provide the stability limits of the closed loop system, including the human operator, the controller and the flexible quadrotor. To the best of authors' knowledge, no similar work exists in the literature, where both the hybrid modeling and the human-in-the-loop stability analysis of a flexible quadrotor UAV, in the presence of an adaptive controller, are presented. The involvement of human operator in the overall analysis is especially important to understand the whole cyber physical human system \citep{anna2020enc,albaba2019,eraslan2019}.

This paper is organized as follows. Section 2 presents the modeling of elastic quadrotor dynamics. The controller design and human-in-the-loop stability analysis are given in Section 3. Simulation results are presented in Section 4 and a summary is given in Section 5.

\section{Modeling of Elastic Quadrotor Dynamics}
In this section, we represent the dynamic modeling of a quadrotor UAV considering elastic effects. In obtaining the nonlinear equations of motion, the Lagrangian method is used \citep{wasz88,rao2007, bau2009,vepa2014}. Below, we first provide the necessary background for the modeling of unconstrained elastic bodies and then develop the flexible UAV model. We mainly follow the method presented by \citet{wasznasa87}. However, unlike \citet{wasznasa87}, our equations of motion includes the damping effects. Furthermore, whereas \citet{wasznasa87} develop a fixed-wing aircraft model, the modeling in this paper is conducted for a quadrotor geometry and loading conditions.  

\subsection{Dynamics of Unconstrained Elastic Bodies}
\begin{figure}[htb]
	\centering
	\includegraphics[scale=0.50]{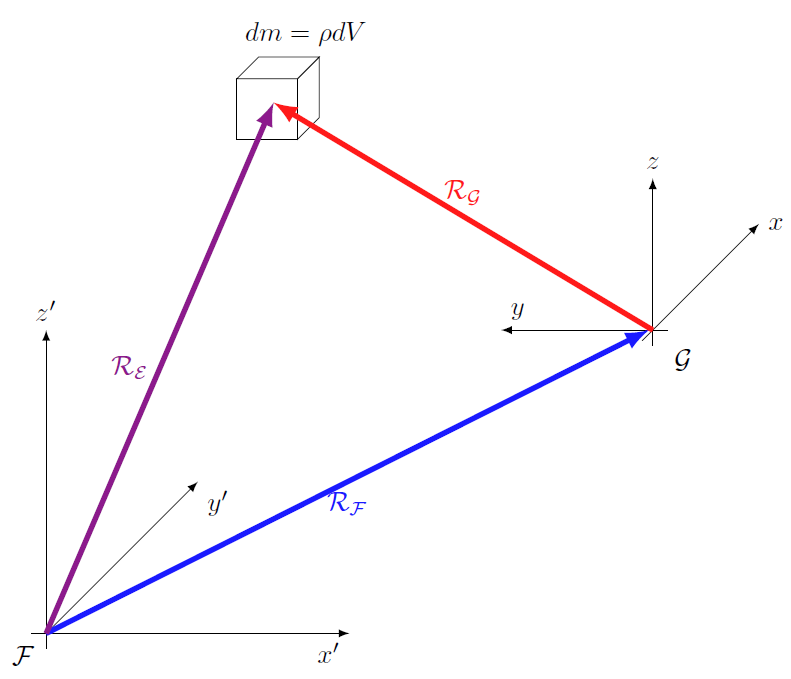}\hspace{5pt}
	\caption{Position of a mass element with respect to reference frames} \label{fig:elasticbodies}
\end{figure}

In an unconstrained elastic body (see Figure \ref{fig:elasticbodies}), the inertial position $\mathcal{R_E}$ of a mass element $\rho dV$, where $\rho$ is the density and $dV$ is the infinitesimal volume, can be obtained by the summation of its position $\mathcal{R_G}$, relative to a non-inertial body-fixed frame $\mathcal{G}$, and the position $\mathcal{R_F}$ of this body reference frame relative to the inertial frame $\mathcal{F}$ as
\begin{equation}\label{eq21}
	\mathcal{R_E}=\mathcal{R_F}+\mathcal{R_G}.
\end{equation}

In the usual rigid body formulation, the time rate of change of $\mathcal{R_G}$ is assumed to be zero \citep{steng2015,vinh95}. This assumption no longer holds true for the elastic body formulation \citep{wasz88}. Denoting $(d/dt)(.)$ as the time derivative with respect to $\mathcal{R_F}$, $(\delta/ \delta t)(.)$ as the time derivative with respect to $\mathcal{R_G}$, and $\omega$ as the angular velocity of $\mathcal{R_G}$ with respect to $\mathcal{R_F}$, the kinetic and potential energy terms, $\mathcal{T}$ and $\mathcal{U}$, respectively, are obtained as \citep{wasznasa87}
\begin{align}\label{energies}
	\begin{split}
			\mathcal{T}&=\frac{1}{2} \int_{V}\left\{\frac{d \mathcal{R_F}}{dt} \cdot \frac{d \mathcal{R_F}}{dt}+2 \frac{d \mathcal{R_F}}{dt} \cdot \frac{\delta \mathcal{R_G}}{\delta t}+\frac{\delta \mathcal{R_G}}{\delta t} \cdot \frac{\delta \mathcal{R_G}}{\delta t}\right.+2 \frac{\delta \mathcal{R_G}}{\delta t} \cdot(\omega \times \mathcal{R_G})\\
			&\left.+(\omega \times \mathcal{R_G}) \cdot(\omega \times \mathcal{R_G})+2(\omega \times \mathcal{R_G}) \cdot \frac{d \mathcal{R_F}}{dt}\right\} \rho dV,\\
	\end{split}
\end{align}
\begin{equation}\label{energiesb}
	\mathcal{U}=-\int_{V}\left(\mathcal{R_F}+\mathcal{R_G}\right) g \rho d V -\dfrac{1}{2} \int_{V} \dfrac{\delta^{2} \mathcal{R_G}}{\delta t^{2}} \mathcal{R_G}  \rho d V.\\
\end{equation}
The position of the mass element $\rho dV$ relative to the body frame $\mathcal{G}$ can be written as
\begin{equation}\label{rgform}
		\mathcal{R_G}= \bar{s}+w(\bar{x},t),  
\end{equation}
where $\bar{s}$ is the constant undeformed length, $w(\bar{x},t)$ is the relative elastic displacement and $\bar{x}$ is the generalized coordinate on the body frame. Assuming that free vibration modes of the elastic body are given, the relative displacement $w(\bar{x},t)$  can be expressed in terms of infinitely many
mode shapes $W(\bar{x})$ and generalized displacement coordinates $\Upsilon(t)$ as
\begin{equation}\label{gendisp}
	w(\bar{x},t)=  \sum_{j=1}^{\infty} W_j(\bar{x})\Upsilon_j(t). 
\end{equation}
Using \eqref{gendisp} and applying the mean axes theorem \citep{wasz88,hesse2014,schmidt98,dussart2018}, \eqref{energies} and \eqref{energiesb} can be rewritten as
\begin{equation}\label{energies-simp}
		\mathcal{T}=\dfrac{1}{2} m \dfrac{d \mathcal{R_F}}{dt} \cdot \dfrac{d \mathcal{R_F}}{dt}+\dfrac{1}{2} \omega^{T}I \omega+\dfrac{1}{2} \sum_{j=1}^{\infty} M_j \dot{\Upsilon}_j^2(t),
\end{equation}
\begin{equation}\label{energies-simp22}		
		\mathcal{U}=-mg\mathcal{R_F} +\dfrac{1}{2} \sum_{j=1}^{\infty} \bar{\omega}_j^2 M_j \Upsilon_j^2(t),   
\end{equation}
where the first, second and the third term in \eqref{energies-simp} are translational, $\mathcal{T}_t$, rotational, $\mathcal{T}_r$ and elastic, $\mathcal{T}_e$, kinetic energy terms, respectively. On the other hand, the first and second term \eqref{energies-simp22} are gravitational, $\mathcal{U}_g$, and elastic, $\mathcal{U}_e$, potential energy terms. The term $M_j$ is the generalized mass term and $\bar{\omega}_j$ is the natural frequency corresponding to the $j^{\mathrm{th}}$ elastic degree of freedom.

\subsection{Equations of Motion for an Elastic Quadrotor UAV}
\begin{figure}[htb]
	\centering
	\includegraphics[scale=0.40]{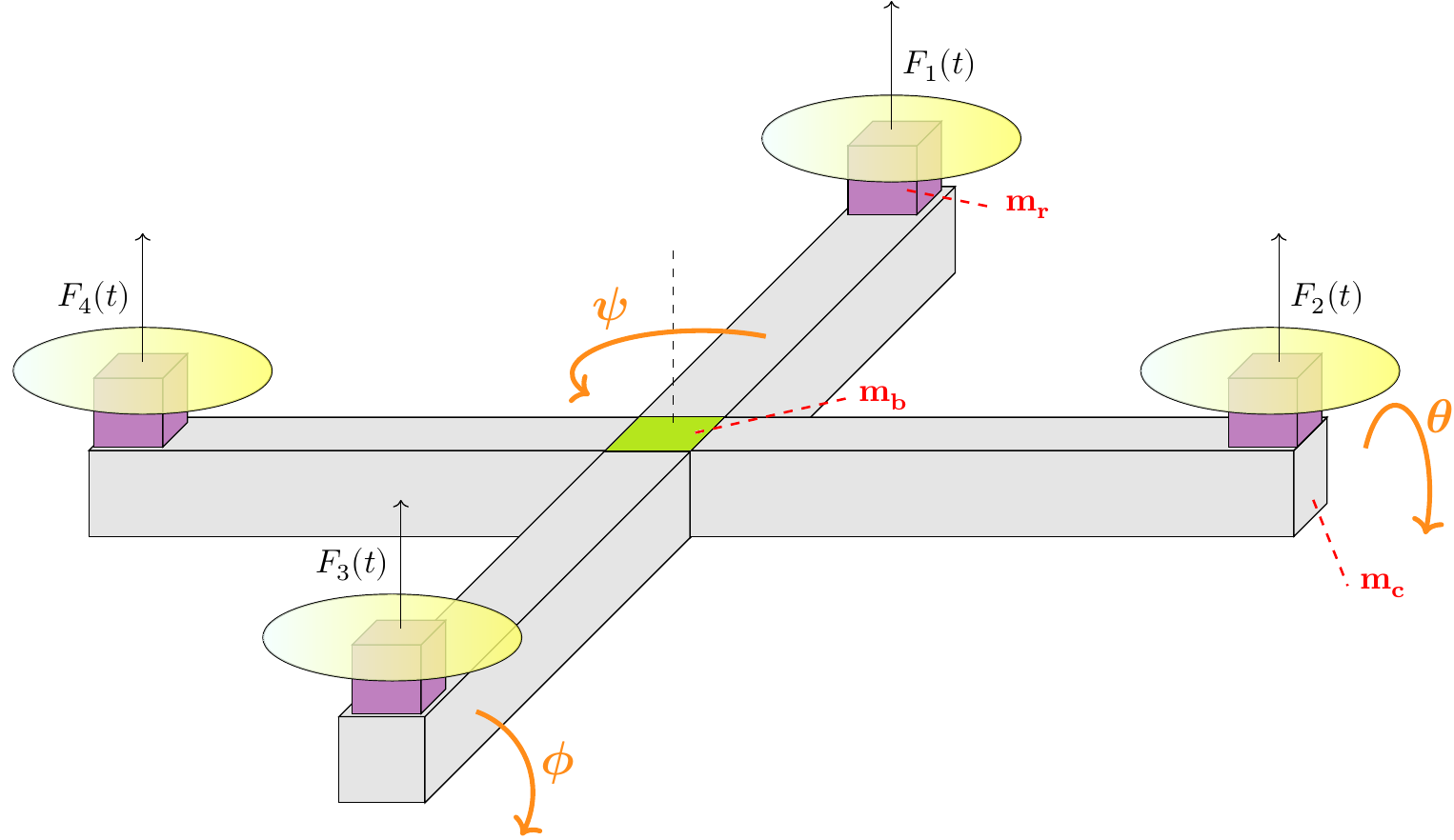}\hspace{5pt}
	\caption{A simplified schematic of the elastic quadrotor UAV} \label{fig:uavbody}
\end{figure}

The elastic quadrotor UAV consists of three different types of masses, that is, the main body mass $m_b$, the arm mass $m_c$ and the rotor mass $m_r$, all of which add up to the total mass $m= m_b+4m_c+4m_r$ (See Figure \ref{fig:uavbody}). The position variable vector and the Euler angles vector pertaining to the center of mass in the body frame are expressed as $\xi = [x, y, z]^T \in \mathbb{R}^3$ and $\upsilon = [\phi, \theta, \psi]^T \in \mathbb{R}^3$, respectively. The inertial angular velocity vector of the center of mass is given by $\omega = [p, q, r]^T \in \mathbb{R}^3$.
The rotation matrix $R^B \in \mathbb{R}^{3 \times 3}$ that transforms the vectors from $\mathcal{G}$ to $\mathcal{F}$ is given as
\begin{equation}\label{rotmatrix}
	R^B=\left
	[\begin{array}{ccc}
		c_\psi c_\theta & c_\psi s_\theta s_\phi-s_\psi c_\phi & c_\psi s_\theta c_\phi + s_\psi s_\phi \\
		s_\psi c_\theta & s_\psi s_\theta s_\phi+c_\psi c_\phi & s_\psi s_\theta c_\phi-c_\psi s_\phi  \\
		-s_\theta & c_\theta s_\phi & c_\theta c_\phi
	\end{array},\right]
\end{equation}
where $s_\upsilon$ and $c_\upsilon$ denote the sine and cosine of the corresponding Euler angle, respectively. The thrust force on the $k^{\mathrm{th}}$ rotor is given by
\begin{equation}\label{thrust}
	F_k = k_t\Omega_k^2,	
\end{equation}
where $k_t$ is the thrust factor and  $\Omega_k$ is the angular velocity of the $k^{\mathrm{th}}$ rotor.
The total thrust force $F_b$ represented in the body frame $\mathcal{G}$ is 
\begin{equation}\label{totalthrust}
	F_b= \sum_{k=1}^{4} k_t\Omega_k^2	
		\left[ 
	\begin{array} { c } 
		0 \\ 
		0 \\ 
		1 
	\end{array}  
	\right] = 
	\left[ 
	\begin{array} { c } 
		0 \\ 
		0 \\ 
		\sum_{k=1}^{4} k_t\Omega_k^2 
	 \end{array}.  
 \right]
\end{equation}
$F_b$ represented in the inertial frame $\mathcal{F}$ is given as
\begin{equation}\label{q_xi}
	\mathcal{Q}_\xi = R^{B} F_b. 
\end{equation}
The torques developed due to the rotational velocities of the rotors are calculated as
\begin{equation}\label{taubody}
	\tau_b = \left[ \begin{array} { c }
		\tau_\phi  
		\\ \tau_ \theta
		\\ \tau_ \psi
	\end{array}
	\right] =  \left[ \begin{array} { c }k_tL_c (\Omega_4^2-\Omega_2^2) \\ k_tL_c (\Omega_3^2-\Omega_1^2) \\ k_q (-\Omega_1^2+\Omega_2^2-\Omega_3^2+\Omega_4^2)  \end{array},  \right]
\end{equation}
where $k_q$ is the drag factor and $L_c$ is the arm length. Gyroscopic torques are given as
\begin{equation}\label{taugyro}
	\begin{aligned}
		\begin{split}
		\tau_g & =  - J_r \left[ 
		\dot{\upsilon} \times \left( 
		\begin{array} { c }
			0\\
			0\\  
			1 
		\end{array}  \right) \right] \Omega_g, \\
		\Omega_g&= \Omega_1-\Omega_2+\Omega_3-\Omega_4,
		\end{split}
	\end{aligned}
\end{equation}
where $J_r$ is the moment of inertia of the rotor and $\Omega_g$ is the gyroscopic velocity. The total torque, $\mathcal{Q}_\upsilon$, represented in the inertial frame $\mathcal{F}$ is obtained as
\begin{equation}\label{q_upsilon}
	\mathcal{Q}_\upsilon = \tau_b + \tau_g.
\end{equation}
The control input vector, $u$, is taken as
\begin{equation}\label{inputmat}
	\begin{aligned}
		\begin{split}
			u&=\left[
			\begin{array}{cccc}
				k_t & k_t  & k_t  & k_t \\
				0 & -k_t  & 0  & k_t\\
				-k_t & 0 & k_t & 0 \\
				-k_q & k_q  & -k_q  & k_q \\
			\end{array}
			\right]
			\left[
			\begin{array}{c}
				\Omega_1^2\\
				\Omega_2^2\\
				\Omega_3^2\\
				\Omega_4^2\\
			\end{array}\right],\\
		&= R^{\Omega_s} \Omega_s,
		\end{split}
	\end{aligned}	
\end{equation}
where $R^{\Omega_s}$ is the corresponding constant transformation matrix, and $\Omega_s = [\Omega_1^2,\Omega_2^2,\Omega_3^2,\Omega_4^2]^T$ is the vector consisting of the squares of rotational velocities. Another useful transformation matrix is the one that converts the force vector $F = [F_1,F_2,F_3,F_4]^T$ into the control input vector $u$. Multiplying $\Omega_s$ with $k_t$ and dividing each element of  $R^{\Omega_s}$ by $k_t$, it follows from \eqref{inputmat} that
\begin{equation}\label{forceeqn}
	\begin{aligned}
		\begin{split}
			u&=\left( \dfrac{1}{k_t} R^{\Omega_s}  \right) F,\\	
			&=R^{F} F,
		\end{split}
	\end{aligned}		
\end{equation}
where $R^{F}$ is the corresponding constant transformation matrix. The generalized coordinates for the elastic body dynamics is given as $q=[\xi^T, \upsilon^T, \Upsilon_j^T]^T \in \mathbb{R}^{(p+6)}$, $j=1, 2, \dotsc, p$, where $p$ is the number of elastic degrees of freedom, which is infinite in theory but can be truncated to a finite number depending on the level of modeling fidelity. The relation between the rotational velocity vector $\omega$ and time rate of change of the Euler angles vector are expressed as
\begin{equation}\label{rotEul}
	 \begin{aligned}
	 	\begin{split}
	 		\omega&=\left
	 		[\begin{array}{ccc}
	 			-s_\theta & 0 & 1 \\
	 			c_\theta s_\psi & c_\psi & 0 \\
	 			c_\theta c_\psi & -s_\psi & 0
	 		\end{array}\right]\dot{\upsilon},\\
 			&= R^{\upsilon}\dot{\upsilon},
	 	\end{split}
	 \end{aligned}
\end{equation}
where  $R^{\upsilon}$ is the corresponding transformation matrix.
Substituting \eqref{rotEul} into \eqref{energies-simp}, it follows that
\begin{equation}\label{energies-simp2}
			\mathcal{T}(q,\dot{q})=\dfrac{1}{2} m \dfrac{d \mathcal{R_F}}{d t} \cdot \dfrac{d \mathcal{R_F}}{d t}
		+\dfrac{1}{2}\dot{\upsilon}^T R{^{\upsilon^T}}IR{^{\upsilon}}\dot{\upsilon}
		+\dfrac{1}{2} \sum_{j=1}^{\infty} M_j \dot{\Upsilon}_j^2(t).	
\end{equation}
The Lagrangian consisting of the set of generalized coordinates for the elastic quadrotor UAV can be expressed as
\begin{equation}\label{finalLagr}
		\mathcal{L}(q,\dot{q}) = \mathcal{T} - \mathcal{U}.
\end{equation}
The friction term is added exogenously to the formulation in terms of a Rayleigh dissipation function \citep[pp. 543-545]{vepa2014} as
\begin{equation}\label{diss}
	\mathcal{D}(\dot{q})=\frac{1}{2}\sum_{j=1}^{\infty} \sigma_c\dot{\Upsilon}_j^2(t), 
\end{equation}
where the term $\sigma_c$ is the damping coefficient term. The Lagrangian equation with a dissipation function and generalized forces is given as
\begin{equation}\label{lagrangian}
	\frac{d}{dt}\left( \frac{\partial \mathcal{L}}{ \partial \dot{q}_i}  \right) -
	\left( \frac{\partial \mathcal{L}}{ \partial q_i}  \right) +
	\left( \frac{\partial \mathcal{D}}{ \partial \dot{q}_i}  \right)
	= \mathcal{Q}_i,
\end{equation}
where $i=1,2, \dotsc ,(p+6)$, where $Q_i$ is the generalized force. Using \eqref{energies-simp2}-\eqref{lagrangian}, the elastic equations of motion can be obtained as
\begin{equation}\label{noneq}
		\ddot{x} =\left(\cos(\psi) \sin(\theta) \cos(\phi) + \sin(\psi) \sin(\phi)\right)\frac{u_1}{m},
	\end{equation}
\begin{equation}\label{noneq1}
\ddot{y} = \left(\sin(\psi) \sin(\theta) \cos(\phi) - \cos(\psi) sin(\phi)\right) \frac{u_1}{m},
\end{equation}
\begin{equation}\label{noneq2}
\ddot{z} =-g+ \left(\cos(\theta) \cos(\phi)\right) \frac{u_1}{m},
\end{equation}
\begin{equation}\label{noneq3}
	\ddot{\phi} =\dot{\theta}\dot{\psi}\left(\dfrac{J_y-J_z}{J_x} \right) - \dfrac{J_r}{J_x}\dot{\theta}\Omega_g+\dfrac{L_c}{J_x}u_2, 
\end{equation}
\begin{equation}\label{noneq4}
	\ddot{\theta} =\dot{\phi}\dot{\psi}\left(\dfrac{J_z-J_x}{J_y} \right) + \dfrac{J_r}{J_y}\dot{\phi}\Omega_g+\dfrac{L_c}{J_y}u_3,
\end{equation}
\begin{equation}\label{noneq5}
	\ddot{\psi} =\dot{\phi}\dot{\theta}\left(\dfrac{J_x-J_y}{J_z} \right) +\dfrac{1}{J_z}u_4,
\end{equation}
\begin{equation}\label{eleq}
 M_j \ddot{\Upsilon}_j(t) + \sigma_c \dot{\Upsilon}_j(t)  + \bar{\omega}_j^2 M_j \Upsilon_j(t) = \mathcal{Q}_{\Upsilon_j}(t).
\end{equation}

\begin{remark}
	The equations of motion comprise a rigid part \eqref{noneq}-\eqref{noneq5} and an elastic part \eqref{eleq}. The rigid part of the equations of motions is identical to those of a rigid quadrotor UAV \citep{bou2007,sab2015}. On the other hand, the elastic part has a form similar to that of an $p$-many mass spring damper systems,  where $p$ is the number of elastic modes.  
\end{remark}

\subsection{Transverse Vibrations of Elastic Arms}
\begin{figure}[htb]
	\centering
	\includegraphics[scale=0.70]{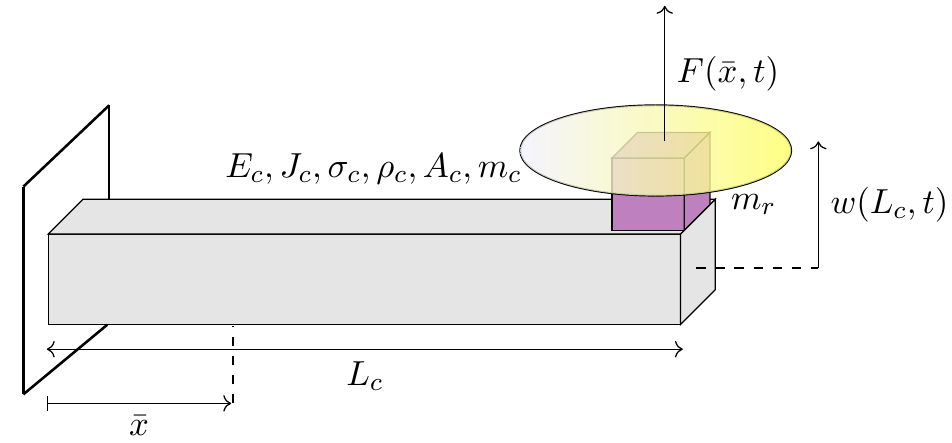}\hspace{5pt}
	\caption{An illustration of the elastic quadrotor arm as a cantilever carrying a rotor. $E_c$, $J_c$, $\sigma_c$ are the Young's modulus, moment of inertia and the damping coefficient of the beam, respectively, $\rho_c$ is the density, $A_c$ is the cross-sectional area, $m_c$ is the mass, $L_c$ is the length of the beam, $m_r$ is the mass of the rotor and $F(\bar{x},t)$ is the concentrated thrust force acting at the beam edge.} \label{fig:cantibeam}
\end{figure}

In the previous subsection, the equations of motion for an elastic quadrotor were derived. The resulting equations of motion for the elastic part \eqref{eleq} are of a relatively simple form, although it is not clear yet what the terms $M_j, \sigma_c, \bar{\omega}_j$ and $\mathcal{Q}_{\Upsilon_j}(t)$ represent in the overall system. In the literature, aeroelastic behavior of flexible aircraft is interpreted as the motion of morphing wings. Upon considering the physical structure of the elastic quadrotor (see Figure \ref{fig:uavbody}), the arms can be modeled as thin cantilever beams undergoing transverse vibrations (see Figure \ref{fig:cantibeam}) owing to continuous motion and agile maneuvers of the quadrotor.

Although a large body of research is devoted to the modeling of undamped Euler-Bernoulli beams under various boundary conditions, relatively small amount of studies can be found for beams with damping: The damping is formulated as an internal property using the viscoelastic Kelvin-Voigt model by \citet{mah2007} and \citet{romas2015}. On the other hand, a model of a cantilever beam with external damping is developed where a dashpot is attached at the free end \citep{fris2001,gurgoze2006}.
For simplicity, we use the latter approach and write the equations of motion governing the damped Euler-Bernoulli beam presented in Figure \ref{fig:cantibeam} as
\begin{equation}\label{pde}
	E_c J_c \frac{\partial ^{4} w(\bar{x},t)}{\partial \bar{x}^{4}} + \rho_c A_c \frac{\partial ^{2} w(\bar{x},t)}{\partial t^{2}} + \sigma_c \frac{\partial  w(\bar{x},t)}{\partial t} = F(\bar{x},t),
\end{equation}
where $E_c$ and $J_c$ are the Young's modulus and moment of inertia of the beam, respectively, $\rho_c$ is the density, $A_c$ is the cross-sectional area, $\sigma_c$ is the damping coefficient of the beam and $F(\bar{x},t)$ is the concentrated thrust force acting at the beam edge. The solution to the homogeneous part of this equation can be obtained by using \eqref{gendisp}, which consists of the mode shape $W_j(\bar{x})$ and the generalized displacement coordinates $\Upsilon_j(t)$. Since the beam is fixed to the moving main rigid body $m_b$ at one end and carries the rotor mass $m_r$ at the other end (see Figure \ref{fig:uavbody}), the boundary conditions can therefore be stated as
\begin{equation}\label{boundaries}
			W(0)=0,
\end{equation}
\begin{equation}\label{boundaries1}
	\frac{dW(0)}{d\bar{x}}=0,
\end{equation}
\begin{equation}\label{boundaries2}
	E_cJ_c\frac{d^2W(L_c)}{d^2\bar{x}}=0,
\end{equation}
\begin{equation}\label{boundaries3}
	E_cJ_c\frac{\partial^3w(L_c,t)}{\partial^3\bar{x}}=m_r\frac{\partial^2w(L_c,t)}{\partial^2 t}.
\end{equation}
Taking $F(\bar{x},t)=0$, substituting \eqref{gendisp} into \eqref{pde}, and solving it together with \eqref{boundaries}-\eqref{boundaries3} \citep{rao2007}, the transcendental frequency equation is obtained as 
\begin{equation}\label{transc}
	1+\frac{1}{\cos \bar{\beta}_j \cosh \bar{\beta}_j}-\bar{m} \bar{\beta}_j(\tan \bar{\beta}_j-\tanh \bar{\beta}_j)=0, 
\end{equation}
\begin{equation}\label{transc1}
    \beta_j=\sqrt[4]{\dfrac{\rho_c A_c \bar{\omega}_j^2}{E_c J_c}},
\end{equation}
where $\bar{\beta}_j=\beta_j L_c$ is the solution of \eqref{transc}, $\beta_j$ is a specific constant obtained from the separation of \eqref{pde} corresponding to the $j^{\mathrm{th}}$ natural frequency $\bar{\omega}_j$, and $\bar{m}=m_r/m_c$ denotes the ratio of the rotor mass $m_r$ to the mass of the cantilever beam $m_c$. For a given $j^{\mathrm{th}}$ mode, we can solve for $\bar{\beta}_j$ in \eqref{transc} and calculate a corresponding natural frequency $\bar{\omega}_j$ in \eqref{transc1}.
Following this procedure, we also obtain the mode shape $W_j(\bar{x})$, which can be written as
\begin{equation}\label{W3}
	W_j(\bar{x})=\bar{\gamma}_j\left[\left(\cos \beta_j \bar{x}-\cosh \beta_j \bar{x}\right)-\frac{\cos \bar{\beta}_j+\cosh \bar{\beta}_j}{\sin \bar{\beta}_j+\sinh \bar{\beta}_j}\left(\sin \beta_j \bar{x}-\sinh \beta_j \bar{x}\right),\right]	
\end{equation}
where $\bar{\gamma}_j$ is a normalization constant corresponding to the $j^{\mathrm{th}}$ mode (See Appendix \ref{app:AppB}). Having found the mode shapes $W_j(\bar{x})$ in \eqref{gendisp}, we are left to find the solutions of the generalized displacement coordinates $\Upsilon_j(t)$ in \eqref{eleq}. Applying orthogonality conditions (see Appendix \ref{app:AppC}), it is obtained that
\begin{equation}\label{modaleq}
	\ddot{\Upsilon}_j(t)+\sigma_c'\dot{\Upsilon}_j(t)+\bar{\omega}_j^2\Upsilon_j(t)=\int_{0}^{L_c}W_j(\bar{x})F(\bar{x},t)d\bar{x},
\end{equation}
where $\sigma_c'=\sigma_c/(\rho_c A_c)$. It is noted that there is a one-to-one correspondence between \eqref{eleq} and \eqref{modaleq}. The generalized mass term $M_j$ in \eqref{eleq} refers to $\rho_c A_c$, which is the mass per unit length of the cantilever beam. Considering the right hand side of \eqref{modaleq} and recalling that $F_k(\bar{x},t)=F_k(t)\delta(\bar{x}-L_c)$ is a concentrated thrust force for the $k^{\mathrm{th}}$ quadrotor arm, $k=\{1,2,3,4\}$, where $\delta(\bar{x})$ is the Dirac's delta function, it can be shown that
\begin{equation}\label{dirac}
	\int_{0}^{L_c}W_j(\bar{x})F_k(t)\delta(\bar{x}-L_c)d\bar{x}=W_j(L_c)F_k(t).
\end{equation}
Substituting \eqref{dirac} into \eqref{modaleq}, we obtain that
\begin{equation}\label{updated}
	\ddot{\Upsilon}_{kj}(t)+\sigma_c'\dot{\Upsilon}_{kj}(t)+\bar{\omega}_{kj}^2\Upsilon_{kj}(t)=W_j(L_c)F_k(t), \quad j=1,2,.., \infty.
\end{equation}
For each arm of the quadrotor, \eqref{updated} has infinitely many solutions corresponding to each $\bar{\omega}_j$. We choose to take the first three natural frequency values, that is, the variable $j$ takes the values of 1, 2 and 3. The relative displacement $w_k(\bar{x},t)$ of the arm $k$ at the tip can then be calculated as
\begin{equation}\label{elstat}
		w_k(L_c,t)=  \sum_{j=1}^{3} W_j(L_c)\Upsilon_{kj}(t).
\end{equation}
Using \eqref{elstat}, we define the corresponding elastic states $z_{kj}$, $k=\{1,2,3,4\}$, $j=\{1,2,3\}$, as
\begin{equation}\label{elstatder}
z_{kj}(t)=W_j(L_c)\Upsilon_{kj}(t),
\end{equation}
\begin{equation}\label{elstatder2}
\dot{z}_{kj}(t)=W_j(L_c)\dot{\Upsilon}_{kj}(t),
\end{equation}
Multiplying \eqref{updated} with $W_j(L_c)$ and using \eqref{elstatder} and \eqref{elstatder2}, \eqref{updated} can be rewritten as
\begin{equation}\label{updated2}
	\ddot{z}_{kj}(t)+\sigma_c'\dot{z}_{kj}(t)+\bar{\omega}_{kj}^2 z_{kj}(t)=W_j^2(L_c)F_k(t).
\end{equation}
This implies that the tip oscillations at each arm $k$ can be modeled
as the summation of solutions of three mass spring damper systems with the same damping coefficient $\sigma_c'$ but different spring constants $\bar{\omega}_{kj}^2$. Therefore, the elastic states for arm $k$ can be written in a state space form as
\begin{equation}\label{statespace1}
	\dot{z}_e^k =A'_e z_e^k+B'_{ze}F_k,
\end{equation}
where $z_e^k=[z_{k1}, \dot{z}_{k1}, z_{k2}, \dot{z}_{k2}, z_{k3}, \dot{z}_{k3}]^T$, and
\begin{equation}\label{statespace2}
	A'_e=\left[
	\begin{array}{cccccc}
		0 & 1 & 0 &0 &0 &0 \\
		-\bar{\omega}_1^2 & -\sigma_c' & 0 &0 &0 &0 \\
		0 & 0 & 0 &1 &0 &0 \\
		0 & 0 & -\bar{\omega}_2^2 &-\sigma_c' &0 &0 \\
		0 & 0 & 0 &0 &0 &1 \\
		0 & 0 & 0 &0 &-\bar{\omega}_3^2 &-\sigma_c'
	\end{array},\right]
	B'_{ze}=
	\left[
		\begin{array}{c}
		0  \\
		W_1^2(L_c)  \\
		0  \\
		W_2^2(L_c)  \\
		0  \\
		W_3^2(L_c)  \\
	\end{array}.\right]
\end{equation}  
Finally, the whole elastic state space formulation can be constructed as
\begin{equation}\label{statespace3}
 	\dot{z}_e =A_e z_e+B_{ze}F,
 \end{equation}
where $F=[F_1, F_2, F_3, F_4]^T$, and
\begin{equation}\label{statespace4}
	A_e=\left[
	\begin{array}{cccc}
		A'_e & 0_{6\times6} & 0_{6\times6} &0_{6\times6}  \\
		0_{6\times6} & A'_e & 0_{6\times6} &0_{6\times6}  \\
		0_{6\times6} & 0_{6\times6} & A'_e &0_{6\times6} \\
		0_{6\times6} & 0_{6\times6} & 0_{6\times6} &A'_e
	\end{array},\right]
	B_{ze}=\left[
	\begin{array}{cccc}
		B'_{ze} & 0_{6\times1} & 0_{6\times1} &0_{6\times1}  \\
		0_{6\times1} & B'_{ze} & 0_{6\times1} &0_{6\times1}  \\
		0_{6\times1} & 0_{6\times1} & B'_{ze} &0_{6\times1} \\
		0_{6\times1} & 0_{6\times1} & 0_{6\times1} &B'_{ze}
	\end{array}.\right]
\end{equation}  
Using \eqref{forceeqn}, the thrust vector can be written in terms of the control input vector $u$ as $F=(R^F)^{-1}u$. Substituting this into \eqref{statespace3}, defining $B_e=B_{ze}(R^F)^{-1}$, and introducing an actuator effectiveness matrix $\Lambda$, it is obtained that
\begin{equation}\label{statespace5}
	\dot{z}_e =A_e z_e+B_{e}\Lambda u.
\end{equation}

\begin{remark}\label{remark2}
	Since the matrix $A_e$ is stable, the subsystem \eqref{statespace5} is bounded-input bounded-states stable. This stability result enables a controller design that is based on rigid body dynamics. However, the designer needs to ensure that 1) control input excitations are not close to the natural frequencies of the elastic modes, and 2) the controller minimizes arm tip oscillations. We discuss these issues in the controller design section below. 
\end{remark}

\section{Controller Design and Human-in-the-Loop Stability Analysis}
\begin{figure}[htb]
	\centering
	\includegraphics[scale=0.70]{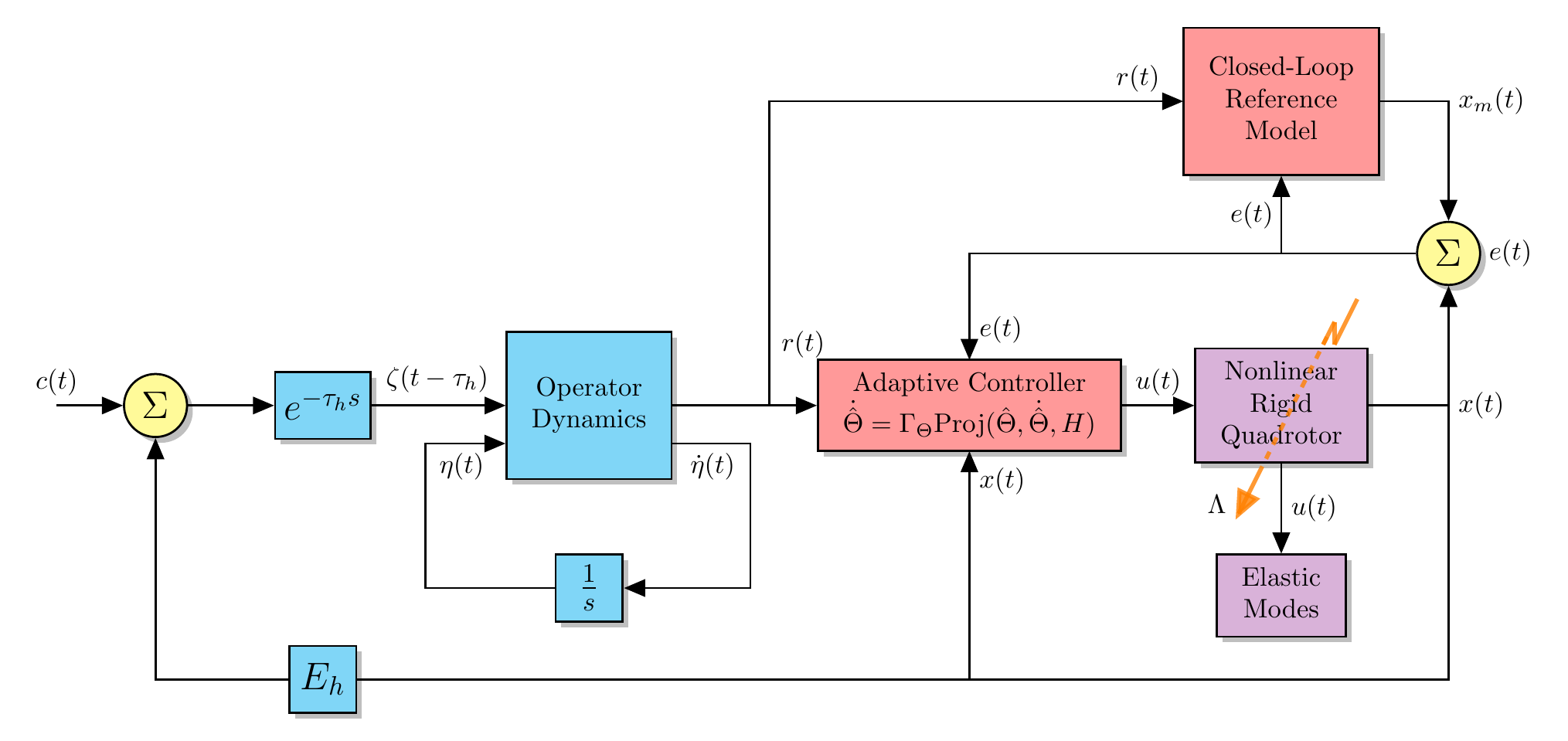}\hspace{5pt}
	\caption{Block diagram of the overall control architecture including the operator dynamics.} \label{fig:blockdiagram}
\end{figure}

The overall closed loop control system consisting of an inner and an outer loop is presented in Figure \ref{fig:blockdiagram}. The inner loop constitutes the uncertain elastic quadrotor dynamics with a closed loop reference model (CRM) adaptive controller. The human operator exists in the outer loop, where s/he observes the commanded and actual plant states, and produces a reference input for the inner loop. Below, we first explain the CRM adaptive controller design and then provide an overall stability analysis in the presence of the human operator. 

\subsection{Controller Design}
Nonlinear equations of motion in \eqref{noneq}-\eqref{noneq5} are linearized around a hover position by performing small angle approximations \citep{dydek2012}. The resulting equations of motion can be represented as

\begin{align}\label{adaptiveplant}
	\begin{split}
		\dot{x}_p(t)&=A_px_p(t)+B_p\Lambda u(t)+B_p\Theta_p^T \Phi_p(x_p(t))\\
		 y_p(t)&=C_p x_p(t),\\
	\end{split}    
\end{align}
where $x_p \in \mathbb{R}^{n_p}$ comprises the position and the Euler angles variables and their corresponding derivatives, $u \in \mathbb{R}^{n_m}$ is the control input, $\Theta_p \in \mathbb{R}^{n_s \times n_m}$ is an unknown weight matrix, $\Phi_p: \mathbb{R}^{n_p} \rightarrow \mathbb{R}^{n_s}$ is a known vector of the form  $ \Phi_{p}(x_p) = [\Phi_{p_1}(x_p),  \Phi_{p_2}(x_p), \dotsm,  \Phi_{p_s}(x_p)]^T$ of high order nonlinear effects and $y_p \in \mathbb{R}^{n_r}$ is the plant output. Besides, $A_p \in \mathbb{R}^{n_p \times n_p}$ is constant and unknown, $B_p \in \mathbb{R}^{n_p \times n_m}$ is a known constant matrix, with the assumption that $(A_p,B_p)$ is controllable, and $\Lambda \in \mathbb{R}^{n_m \times n_m}$ is an unknown positive definite matrix representing the control effectiveness. The evolution of the elastic states is given in \eqref{statespace5}. The control goal of interest is bounded command tracking in the presence of uncertainties, that is, tracking a reference $r(t) \in \mathbb{R}^{n_r}$ produced by the human pilot (See Figure \ref{fig:blockdiagram}). To achieve tracking, a new state vector $e_p \in \mathbb{R}^{n_r} $ is defined as the integral of the tracking error,
\begin{equation}\label{intstate}
e_p(t)= \int_{0}^{t} [y_p(\varepsilon)-r(\varepsilon)] d\varepsilon,	
\end{equation}
and augmented with \eqref{adaptiveplant}, which results in the dynamics
\begin{equation}\label{newplant}
	\dot{x}(t)=Ax(t)+B\Lambda u(t) + B\Theta_p^T \Phi_p(x_p(t)) + B_m r(t),
\end{equation}
where
\begin{equation}\label{newplantconstants}
		A =  \begin{bmatrix}
			A_p & 0_{n_p \times n_m} \\
			C_p  & 0_{n_m \times n_m}  \\
		\end{bmatrix}, \enspace
		B =  \begin{bmatrix} 
			B_p  \\
			0_{n_m \times n_m}   
		\end{bmatrix}, \enspace \\
		B_m =  \begin{bmatrix}
			0_{n_p \times n_r}  \\
			-I_{n_r \times n_r}   \\
		\end{bmatrix},
\end{equation}
and $x(t)=[x_p(t)^T, e_p(t)^T]^T \in \mathbb{R}^{(n_p+n_m)}$ is the augmented state vector with $n=n_p+n_m$. The control law is determined as
\begin{equation}\label{inputvalues}
	u(t)= u_{\mathrm{bl}}(t)+u_{\mathrm{ad}}(t),
\end{equation}
where $u_{\mathrm{bl}}(t) \in \mathbb{R}^{n_m} $ and $u_{\mathrm{ad}}(t) \in \mathbb{R}^{n_m}$ are the baseline and the adaptive control laws, respectively. The baseline controller is given as
\begin{equation}\label{ubl}
	u_{\mathrm{bl}}(t)=-K^Tx(t),
\end{equation}
where $K \in \mathbb{R}^{n \times n_m}$ is a fixed state feedback control gain matrix. We choose this gain such that
\begin{equation}\label{refam}
	A_m= A-B\Lambda K^T
\end{equation}
becomes a stable matrix. The reference model is selected as
\begin{equation}\label{reff}
	\dot{x}_m(t)=A_m x_m(t) + B_m r(t)-Le(t),
\end{equation} 
where $x_m \in \mathbb{R}^{n} $ is the reference model state vector, $e(t)=x(t)-x_m(t)$ is the tracking error and $L \in \mathbb{R}^{n \times n}< 0$ is a constant matrix such that $(A_m+L)$ is Hurwitz. Substituting \eqref{inputvalues}, \eqref{ubl} and \eqref{refam} into \eqref{newplant}, one obtains
\begin{equation}\label{reflikeplant}
	\dot{x}(t)=A_mx(t) + B_m r(t) + B\Lambda[u_{\mathrm{ad}}(t)+ \Theta^T \Phi(x(t))],
\end{equation}
where $ \Theta^T= \left[ \Lambda^{-1}\Theta_p^T \right] \in \mathbb{R}^{n_m\times n_s}$ is the unknown overall weight matrix and $\Phi^T(x(t))=\left[\Phi_p^T(x_p(t)) \right] \in \mathbb{R}^{(n_s+n)}$ is a vector of high order nonlinear effects. We choose an adaptive control of the form
\begin{equation}\label{uad}
	u_{\mathrm{ad}}(t)=-\hat{\Theta}^T\Phi(x(t)),
\end{equation}
where $\hat{\Theta} \in \mathbb{R}^{(n_s+n)\times n_m}$ is the matrix of time-varying adaptive parameters. The adaptive law is given by
\begin{equation}\label{uadlaw}
	\dot{\hat{\Theta}}=\Gamma_{\Theta} \Phi(x(t))e^T(t)PB,
\end{equation}
where $\Gamma_{\Theta} \in \mathbb{R}^{[(n_s+n)\times n_m]\times [(n_s+n)\times n_m]}$ is a diagonal positive definite matrix of
adaptive gains and $P \in \mathbb{R}^{n \times n}$ is the unique symmetric positive definite solution of the Lyapunov equation
\begin{equation}\label{lyapeq}
	(A_m+L)^T P + P (A_m+L) = -Q,
\end{equation}
where $Q \in \mathbb{R}^{n \times n}>0$ is a positive definite symmetric matrix. To prevent adaptive parameter drifts, the projection algorithm \citep{gibson2012,tran2018,tohidi2020,islam2018,laf2020} is employed as
\begin{equation}\label{projj}
	\dot{\hat{\Theta}}=\Gamma_{\Theta} \operatorname{Proj(\hat{\Theta}, \Phi(x(t))e^T(t)PB,H)},
\end{equation} 
where the projection operator is defined as
	\begin{equation}\label{projdef1}
	\operatorname{Proj}(\Theta, Y, H)=\left[\operatorname{Proj}\left(\theta_{1}, y_{1}, h_{1}\right) \ldots \operatorname{Proj}\left(\theta_{m}, y_{m}, h_{m}\right)\right],
	\end{equation}
	where $\Theta=\left[\theta_{1} \ldots \theta_{m}\right] \in \mathbb{R}^{(n_s+n) \times n_m}$, $Y=\left[y_{1} \ldots y_{m}\right] \in \mathbb{R}^{(n_s+n) \times n_m}$, and
	 $H=\left[h_{1}\left(\theta_{1}\right) \ldots h_{m}\left(\theta_{m}\right)\right]^{T} \in$
	$\mathbb{R}^{n_m \times 1}$. The vector form of the projection operator is 
	\begin{equation}\label{projy}
		\operatorname{Proj}\left(\theta_{j}, y_{j}, h_{j}\right)=\left\{\begin{array}{ll}
			y_{j}-\frac{\nabla h_{j}\left(\theta_{j}\right)\left(\nabla h_{j}\left(\theta_{j}\right)\right)^{T}}{\left\|\nabla h_{j}\left(\theta_{j}\right)\right\|^{2}} y_{j} h_{j}\left(\theta_{j}\right) & \text { if } h_{j}\left(\theta_{j}\right)>0 \wedge y_{j}^{T} \nabla h_{j}\left(\theta_{j}\right)>0 \\
			y_{j} & \mathrm{otherwise}
	\end{array}\right.
	\end{equation}
where $h: \mathbb{R}^{n_m} \rightarrow \mathbb{R}$ is a convex function and $\nabla h(\theta)=\left(\frac{\partial h(\theta)}{\partial \theta_{1}} \cdots \frac{\partial h(\theta)}{\partial \theta_{n_m}}\right)^{T}$. Defining the adaptive parameter estimation error as $\tilde{\Theta}=\hat{\Theta}-\Theta$, and subtracting \eqref{reff} from \eqref{reflikeplant}, the reference model tracking error can be obtained as  
\begin{equation}\label{errordyn}
	\dot{e}(t)=A_m e(t)-B \Lambda \tilde{\Theta}^T\Phi(x(t))+Le(t).
\end{equation}
Using the Lyapunov function candidate
\begin{equation}\label{lyappa}
	\mathcal{V}(e,\tilde{\Theta})=e^T(t)Pe(t) + \mathrm{tr}[(\tilde{\Theta}^T\Gamma^{-1}\tilde{\Theta})\Lambda],
\end{equation}
it can be shown that 
\begin{equation}\label{lyapder}
\dot{\mathcal{V}}(e(t), \tilde{\Theta}(t))=-e^{\mathrm{T}}(t) \operatorname{Pe}(t) \leq 0.	
\end{equation}
This implies that the equilibrium point of \eqref{uadlaw} and \eqref{errordyn} is stable in the sense of Lyapunov. The convergence of $e$ to zero can typically be shown using Barbalat's Lemma. However, here the lemma is inapplicable, since $\ddot{\mathcal{V}}(e(t), \tilde{\Theta}(t))$ cannot be proven to be bounded, yet. The term $x(t)=e(t)+x_m(t)$ contains the reference model state $x_m(t)$, which can grow unboundedly due to the reference $r(t)$ produced by the human pilot model. For this reason, the dynamics of the outer loop needs to be investigated to determine  whether or not $x_m(t)$ and $r(t)$ are bounded.
\subsection{Outer Loop Dynamics}
We use a linear model with a time delay for human operator dynamics, represented as
\begin{equation}\label{behuman}
		\dot{\eta}(t)=A_{h} \eta(t)+B_{h} \zeta(t-\tau_h),
\end{equation}
\begin{equation}\label{behuman2}
	r(t)=C_{h} \eta(t)+D_{h} \zeta(t-\tau_h),
\end{equation}
where $\eta(t) \in \mathbb{R}^{n_{\eta}}$ is the human state vector, $\tau_h \in$ $\mathbb{R}^{+}$ is the reaction delay, and $A_{h} \in \mathbb{R}^{n_{\eta} \times n_{\eta}}, B_{h} \in$
$\mathbb{R}^{n_{\eta} \times n_{c}}, C_{h} \in \mathbb{R}^{n_{r} \times n_{\eta}}$, and $D_{h} \in \mathbb{R}^{n_{r} \times n_{c}}$ are constant matrices. $r(t) \in \mathbb{R}^{n_{r}}$ is the
reference formed by the human operator (see Figure \ref{fig:blockdiagram}). The input to the human dynamics is a feedback error term of the form
\begin{equation}\label{humaninput}
	\zeta(t) = c(t)-E_{\mathrm{h}} x(t),
\end{equation}
where $E_h \in \mathbb{R}^{n_c \times n}$ is a constant matrix that allows to choose a subset of the state $x(t)$ as feedback. Similar human models, containing a linear part and a time delay can also be found in \citep{thur2000,witt2004,mill2011}. The analysis in this chapter follows the similar steps used in \citet{yucelen2017}. Using \eqref{behuman2} and \eqref{humaninput}, \eqref{reff} and \eqref{behuman} can be rewritten as 
\begin{equation}\label{refhuman}
	\begin{aligned}
		\begin{split}
			\dot{x}_{{m}}(t)
		=& A_{{m}} x_{{m}}(t)+B_{{m}} C_{{h}} \eta(t) -B_{{m}} D_{{h}} E_{{h}}[x_{{m}}(t-\tau_h)+e(t-\tau_h)]\\
		+&B_{{m}} D_{{h}} c(t-\tau_h)-Le(t),
		\end{split}
	\end{aligned}
\end{equation}
\begin{equation}\label{refhuman2}
			\dot{\eta}(t)=A_{{h}} \eta(t)-B_{{h}} E_{{h}} x_{{m}}(t-\tau_h) -B_{{h}} E_{{h}} e(t-\tau_h)
			+B_{{h}} c(t-\tau_h).
\end{equation}
Defining $\mu(t) \triangleq \left[x_{{m}}^{{T}}(t), \eta^{{T}}(t)\right]^{{T}}$, \eqref{refhuman} and \eqref{refhuman2}  can be represented as a single delay equation as
\begin{equation}\label{dde}
	\dot{\mu}(t)=\mathcal{A}_{n} \mu(t)+\mathcal{A}_{d} \mu(t-\tau_h)+\Pi(\cdot),
\end{equation}
where
\begin{equation}\label{AnAd}
	\begin{aligned}
		\begin{split}
			\mathcal{A}_{n} &=\left[
			\begin{array}{cc}
			A_{{m}} & B_{{m}} {C}_{{h}} \\
			0_{n_{\eta} \times n} & A_{{h}}
			\end{array},\right]\\
		\mathcal{A}_{d} &=\left[
		\begin{array}{cc}
			-B_{{m}} D_{{h}} E_{{h}} &  0_{n \times n_{\eta}} \\
			-B_{{h}} E_{{h}} &  0_{n_{\eta} \times n_{\eta}}
		\end{array},\right]\\
	\Pi(\cdot) &=\left[
	\begin{array}{c}
		-Le(t)-B_{{m}} D_{{h}} E_{{h}} e(t-\tau_h)+B_{{m}} D_{{h}} c(t-\tau_h) \\
		-B_{{h}} E_{{h}} e(t-\tau_h)+B_{{h}} c(t-\tau_h)
	\end{array}.\right]
		\end{split}
	\end{aligned}
\end{equation}
Since $e(t)$ is shown to be bounded in the previous section and the command $c(t)$ is assumed to be bounded, the matrix $\Pi(\cdot)$ is bounded.
\begin{theorem}\label{theo31}
	Consider the dynamics given in \eqref{dde}. If the real parts of all the infinitely many roots of the equation
\end{theorem}
\begin{equation}\label{deter}
\operatorname{det}\left(s I-\left(\mathcal{A}_{n}+\mathcal{A}_{d} e^{-\tau_h s}\right)\right)=0	
\end{equation}
have strictly negative real parts, then $\mu(t) \in \mathcal{L}_{\infty}$ and $\lim _{t \rightarrow \infty} e(t)=0$.
\begin{proof}
	If all of the roots of the characteristic equation \eqref{deter} have strictly negative real parts, then the homogeneous part of \eqref{dde}, given as
	\begin{equation}\label{ddeh}
		\dot{\mu}(t)=\mathcal{A}_{n} \mu(t)+\mathcal{A}_{d} \mu(t-\tau_h)
	\end{equation}
	is stable. Furthermore, since the forcing term $\Pi(\cdot)$ in \eqref{dde} is bounded, the the solution $\mu(t)$ is bounded. This implies that both the reference model state $x_m(t)$ and the human state $\eta(t)$ are bounded. At this point, it can be shown that the second derivative of the Lyapunov function \eqref{lyappa} is bounded. Hence, with the application of Barbalat's Lemma it can be shown that $\lim _{t \rightarrow \infty} e(t)=0$. 
\end{proof}
\begin{remark}
Depending on the application, stability limits of the overall system change based on  the roots of \eqref{deter}. In the following section, we conduct this analysis for our simulation example. 
\end{remark}

\section{Simulations}
In this section, a number of simulations are performed in order to demonstrate the stability and performance characteristics of the human-in-the-loop control system, consisting of the human operator, the controller and the flexible UAV. Below, we first explain the simulation scenario, the controller design details, the stability limits of the operator dynamics and then discuss the simulation results. 

\subsection{Simulation Scenario}  
In the simulations, the elastic UAV equations of motion introduced in \eqref{noneq}-\eqref{noneq5} and \eqref{statespace5} are used as the plant model. The human operator is assumed to behave like a proportional integral (PI) controller, with a reaction time delay. This model is consistent with the operator model introduced in \eqref{behuman}-\eqref{humaninput}, and can be represented as 
\begin{equation}\label{humanpilot1EE}
	G_{\mathrm{PI}}(s)=K_p \frac{T_ps+1}{s}e^{-\tau_h s},
\end{equation}
where $K_p>0$ and $T_p>0$ are model constants, and $\tau_h$ is the human operator reaction time delay. The parameters used in the UAV and operator models are given in Table \ref{tab:datatableEE}. 

Two types of flight conditions are simulated: operator controlled and autonomous flight. In the operator controlled flight, the human operator's goal is to make the UAV follow a desired altitude command $z_d$, by producing a corresponding reference input, which is fed to the controller (See Figure \ref{fig:blockdiagram}). During this flight mode, the rest of the position and attitude references, $x_d$, $y_d$ and $\psi_d$, are created externally. In the autonomous flight mode, all of the reference inputs are created externally and achieved autonomously by the controller, without any interference from the operator. In a simulation of 70 seconds, two anomalies are injected at $t_a=16s$, which result in loss of control effectiveness of $75 \%$ and $50 \%$ in the second and third rotors, respectively.

\begin{table}[htb]
	\tbl{Elastic UAV Model and Human Operator Parameters}
	{\begin{tabular}{lccccccc} \toprule
			
			Quadrotor Body& Value & Unit & Arms & Value & Unit & Operator & Value \\ \midrule
			$m$ & 0.5 & $kg$  & $L_c$ & 0.21 & $m$ & $K_p$ & 0.59\\
			$J_x$ & $4.85 \times 10^{-3}$ & $kgm^2$ & $\rho_c$ & 1370 & $kgm^3$ & $T_p$ & 0.41 \\
			$J_y$ & $4.85 \times 10^{-3}$ & $kgm^2$ & $E_c$ & 2.91 & $GPa$ & $\tau_h$ & 0.20 \\
			$J_z$ & $8.81 \times 10^{-3}$ & $kgm^2$ &  &  &  & & \\
			$J_r$ & $3.36 \times 10^{-5}$ & $kgm^2$  &  &  &  & & \\ 
			\bottomrule
	\end{tabular}}
	%\tabnote{\textsuperscript{a}This footnote shows how to include
	%	footnotes to a table if required.}
	\label{tab:datatableEE}
\end{table}

\subsection{Controller Design Details} 
The baseline controller gain vector $K$ is first calculated based on the nominal plant dynamics. Then, the elements of this vector is decreased by $20\%$ to introduce additional uncertainty. For the design of the adaptive controller, three sets of design parameters need to be determined: Adaptation rates, initial adaptive parameter values and projection boundaries. An empirical approach that assumes that the control parameters reach their ideal values within three time constants is employed to determine the adaptation rates.  \citep{dydek2006,yildiz2010}. This method can mathematically be expressed as
\begin{equation}\label{adspeedEE}
	\Gamma_{\Theta_{ii}} = \dfrac{\|\Theta_i\|}{3\tau_m |\bar{r}^2|},
\end{equation}
where $\tau_m$ is the smallest time constant of the reference model $A_m$ and $\bar{r}$ is the maximum value of the reference. Since the ideal control parameter values are unknown, the nominal ideal values (calculated using the nominal plant dynamics) are used instead. It is noted that \eqref{adspeedEE} is mainly used as a starting point for fine-tuning the adaptation rates. The initial conditions of all the adaptive control parameters are set to zero. Finally, the projection boundaries are selected by observing the variation of controller parameters during simulations.

\subsection{Stability Limits}
\begin{figure}[htb]
	\centering
	\includegraphics[scale=0.25]{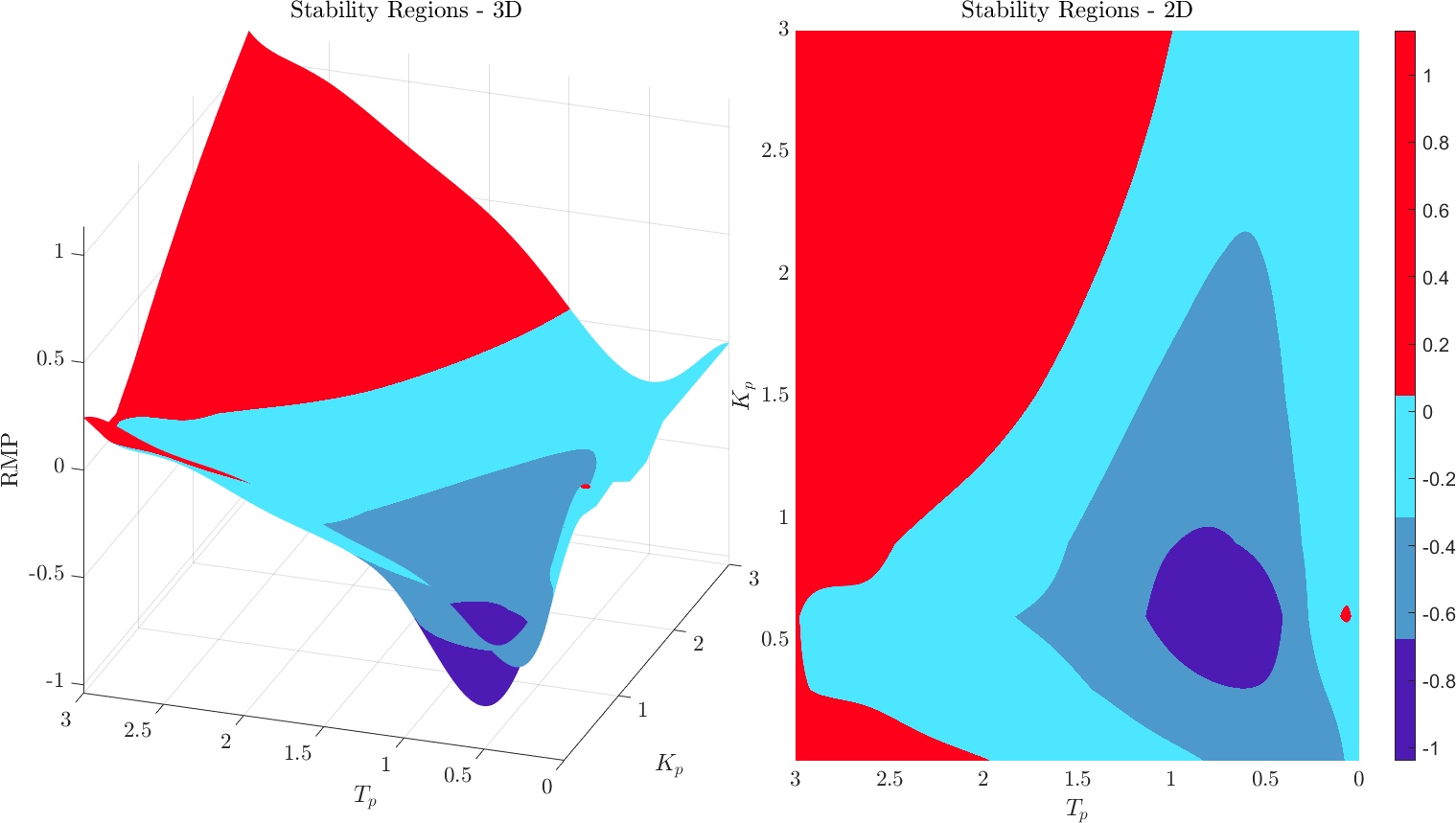}\hspace{5pt}
	\caption{Variation of rightmost pole location with respect to simultaneous change in $K_p$ and/or $T_p$.} \label{fig:Surf3D}
\end{figure}

As stated in Theorem \ref{theo31}, once the CRM adaptive controller is designed as given in \eqref{inputvalues}-\eqref{projy}, the stability of the overall system is determined by the roots of the characteristic polynomial presented in \eqref{deter}. We use the DDE-BIFTOOL \citep{engelborghs2012} to find the rightmost root, among infinitely many of them, of this polynomial for the simulation example. Specifically, we are interested in the effect of the operator parameters $K_p$ and $T_p$ in \eqref{humanpilot1EE} on the stability of the overall system. Figure \ref{fig:Surf3D} shows the location of the rightmost root of the characteristic polynomial \eqref{deter} for different values of $K_p$ and $T_p$. The red areas in the figure represent the unstable regions. It can be argued that the system can be swept into the unstable region for moderately high values of $T_p$. In addition, the relatively small patch of instability around $K_p=0.6$ and $T_p=0.07$ shows the possibility of unexpected system behavior due to operator time-delays. 

\subsection{Simulation Results}
\begin{figure}[htb]
	\centering
	\includegraphics[scale=0.25]{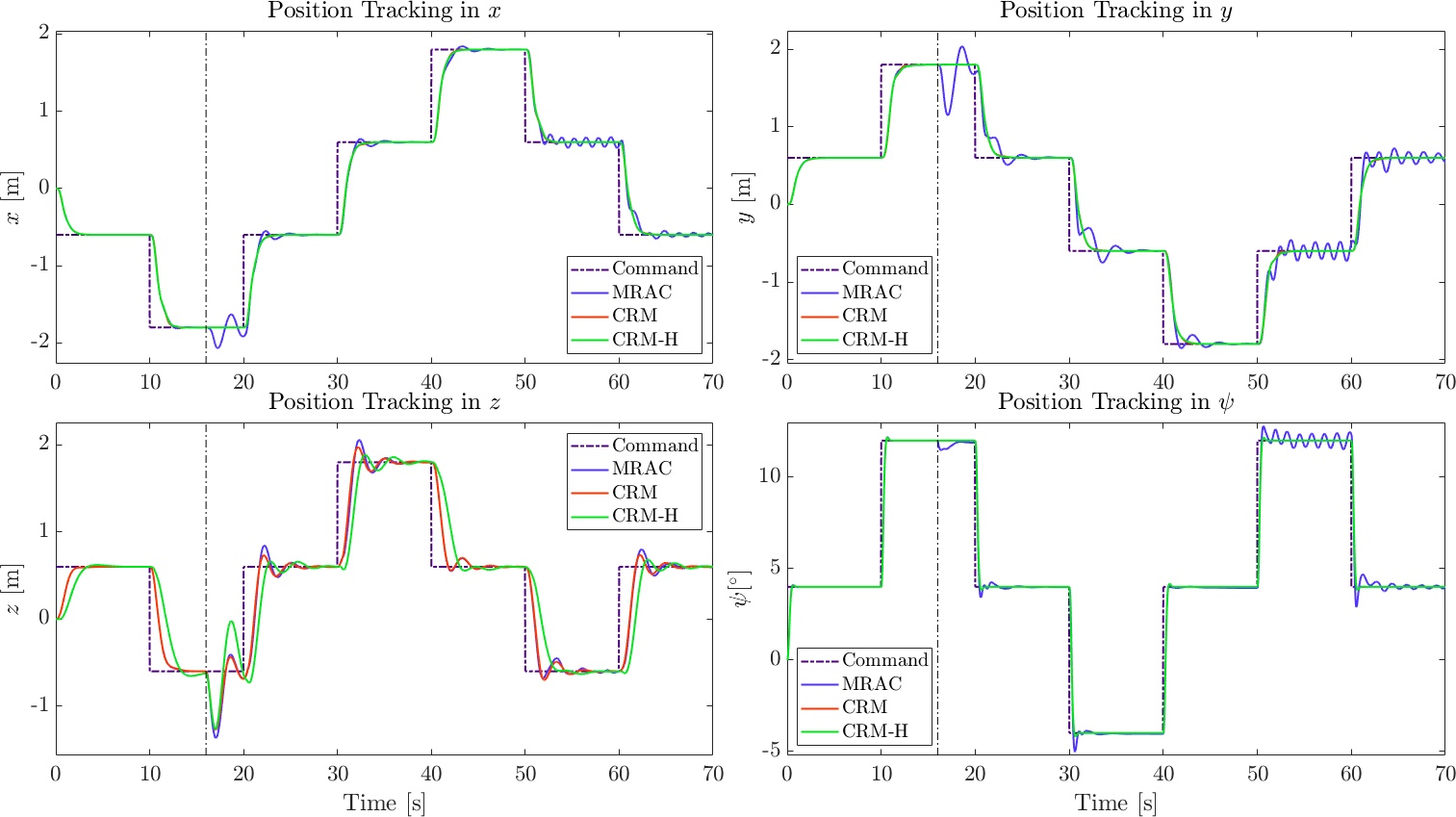}\hspace{5pt}
	\caption{The position tracking performance of the MRAC, CRM and CRM-H configurations.} \label{fig:P1EE}
\end{figure}

\begin{figure}[htb]
	\centering
	\includegraphics[scale=0.25]{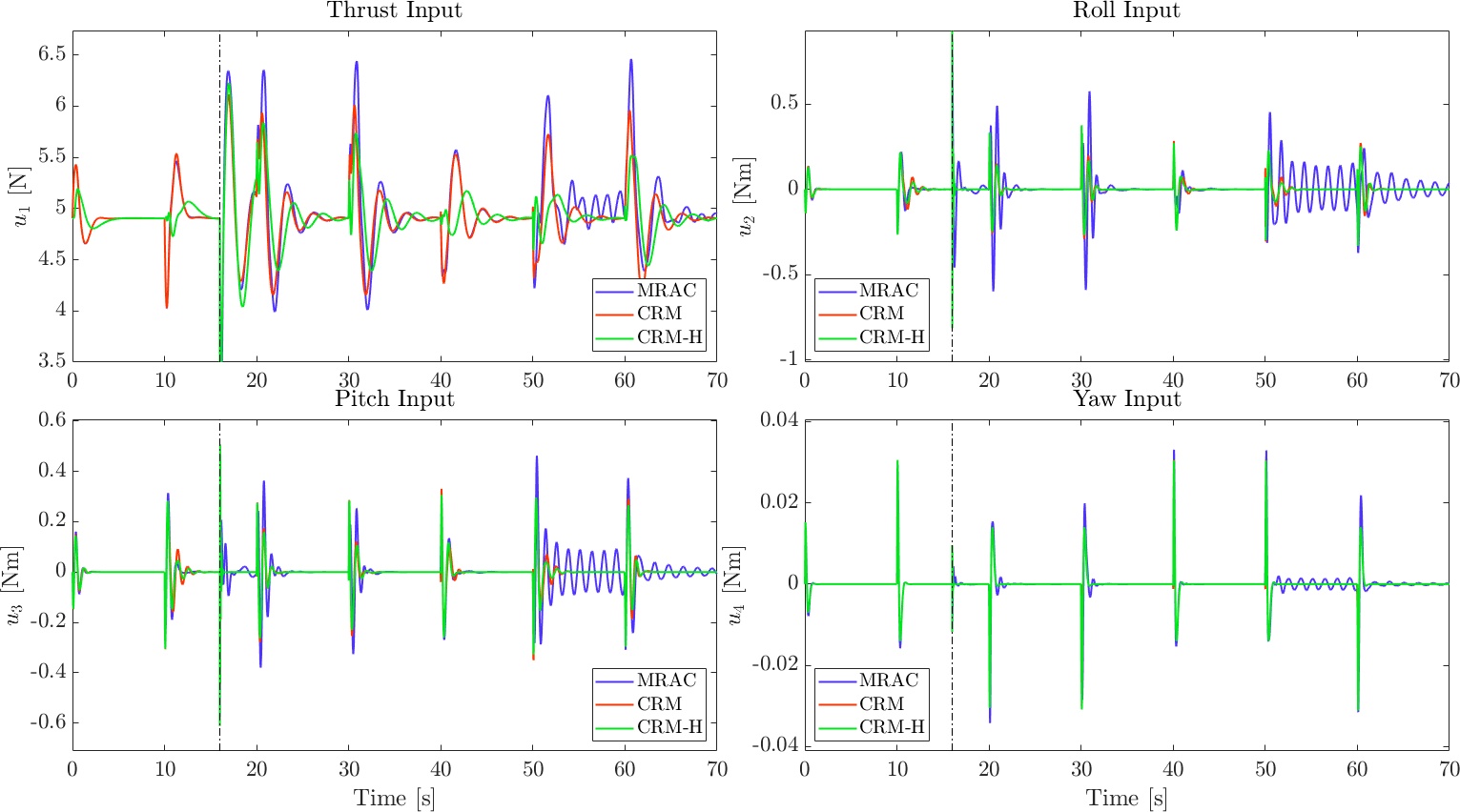}\hspace{5pt}
	\caption{The control inputs of the MRAC, CRM and CRM-H configurations.} \label{fig:P2EE}
\end{figure}

Tracking performances of three different closed loop control systems are presented in Figure \ref{fig:P1EE}. In the figure, the autonomous flights using a model reference adaptive controller and a closed loop reference model adaptive controller are labeled as MRAC, and CRM, respectively. Human operator controlled flight, where a CRM is used as the controller (See Figure \ref{fig:blockdiagram}) is labeled as CRM-H. This figure, together with Figure \ref{fig:P2EE} show that CRM based configurations induce smoother trajectory responses and control inputs. The effect of human operator involvement is also observed as delayed responses to commanded inputs, due to human reaction lag. Overall performance of different configurations, which is defined by the metric
\begin{equation}\label{rms-errorEE}
	\mathcal{M}_e= \mathrm{rms}(e(t))|_{t_a}^{T}, \quad
\end{equation}
where $e(t)$ is the reference model tracking error is provided in Table \ref{tab:mraccrmHH}. As expected, CRM based controllers provide better tracking performances. 
	\begin{table}[htb]
	\tbl{The tracking performance assessment metric for the MRAC, CRM and CRM-H configurations.}
	{\begin{tabular}{lccc} \toprule
			Axes & $\mathcal{M}{_e}^{\mathrm{MRAC}}$ & $\mathcal{M}{_e}^{\mathrm{CRM}}$ & $\mathcal{M}{_e}^{\mathrm{CRM-H}}$   \\ \midrule
			$x$ & 5.153 & 0.039  & 0.032  \\
			$y$ & 11.544 & 0.041 & 0.029  \\
			$z$ & 13.138 & 3.436 & 3.435 \\
			$\psi$ & 0.378 & 0.001 & 0.001  \\
			\bottomrule
	\end{tabular}}
	%\tabnote{\textsuperscript{a}This footnote shows how to include
	%	footnotes to a table if required.}
	\label{tab:mraccrmHH}
\end{table}

\begin{figure}[htb]
	\centering
	\includegraphics[scale=0.25]{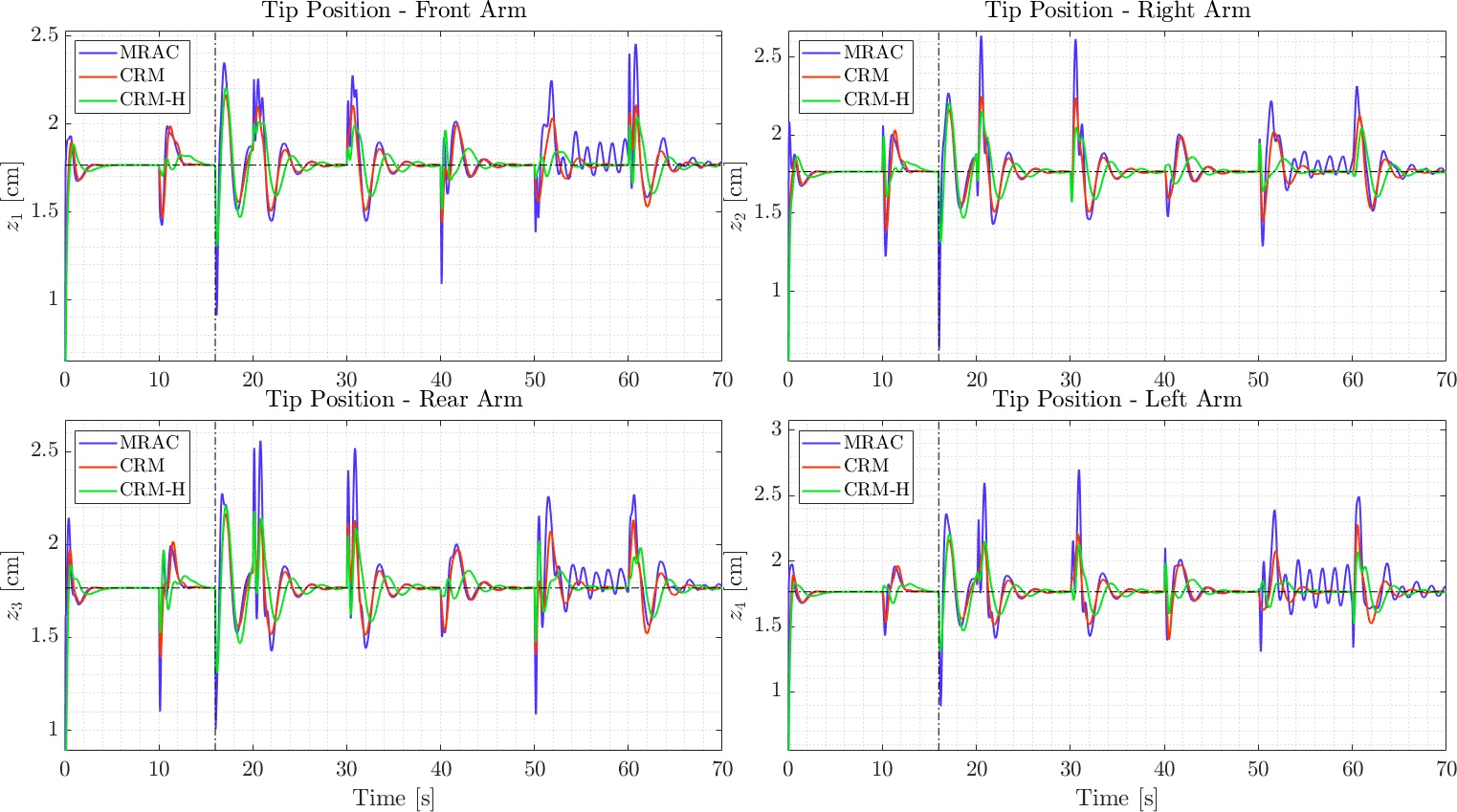}\hspace{5pt}
	\caption{The arm tip oscillations of the MRAC, CRM and CRM-H configurations.} \label{fig:P8EE}
\end{figure}

\begin{figure}[htb]
	\centering
	\includegraphics[scale=0.25]{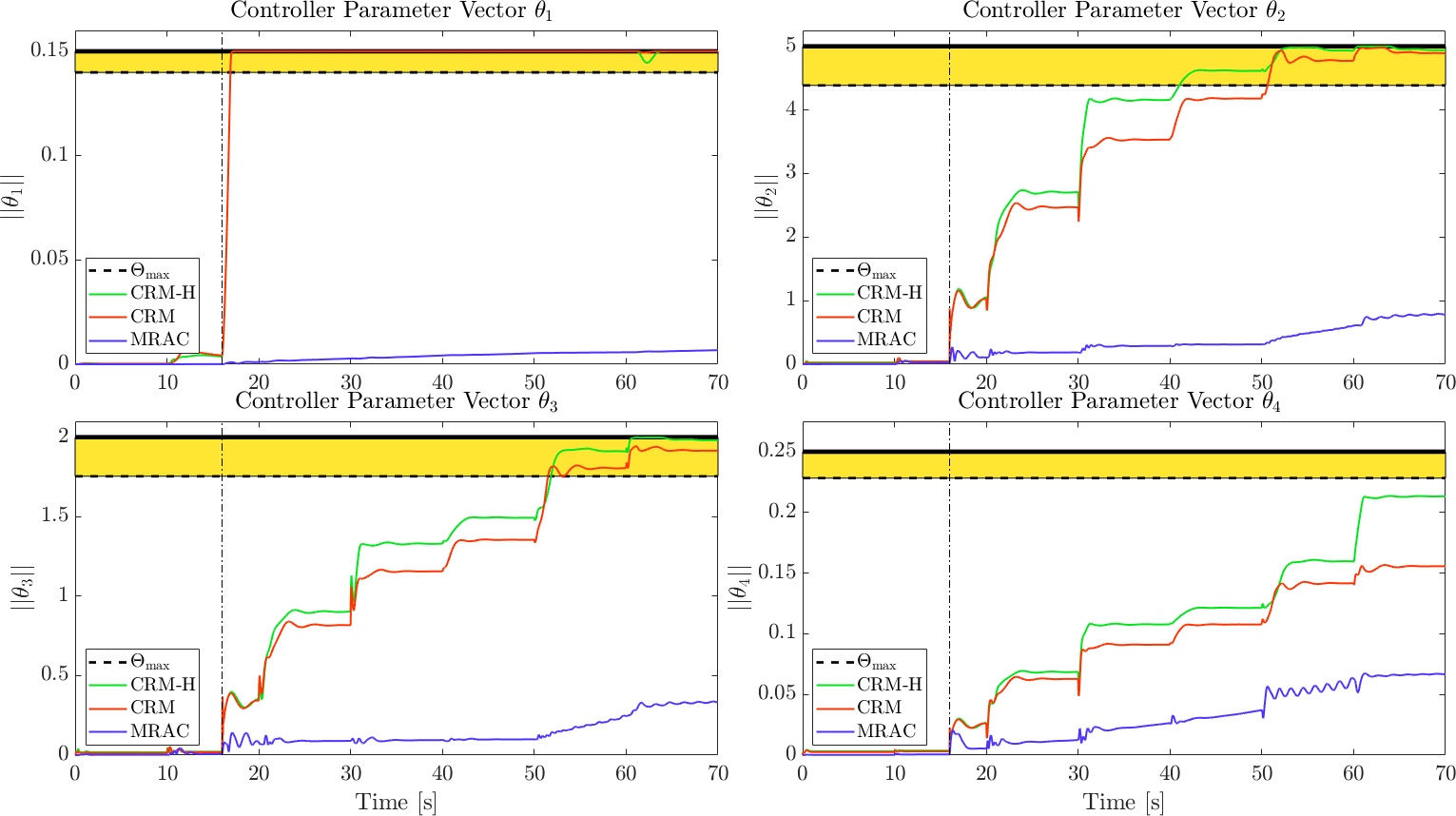}\hspace{5pt}
	\caption{The evolution of control parameters of the MRAC, CRM and CRM-H configurations.} \label{fig:P4EE}
\end{figure}

As previously stated in Remark \ref{remark2}, it should be ascertained whether or not control input excitations are close to the natural frequencies of the elastic modes. The natural frequencies, $\bar{\omega}_{kj}$, $k=\{1,2,3,4\}$, $j=\{1,2,3\}$, of the first three elastic modes of four quadrotor arms (see \eqref{updated2}) are calculated as $131$ $\mathrm{rad/s}$, 1365 $\mathrm{rad/s}$ and 1865 $\mathrm{rad/s}$, respectively. Figure \ref{fig:P2EE} demonstrates that none of the controllers excite these frequencies. On the other hand, it is shown in Figure \ref{fig:P8EE} that the arm tip oscillations are lowest in CRM based configurations, which could be predicted from the quadrotor trajectories provided in Figure \ref{fig:P1EE}.  
Variation of the adaptive parameters under the effect of reference changes and anomalies is presented in Figure \ref{fig:P4EE}. The horizontal dashed black lines in the subfigures denote the projection boundaries. Yellow bands are projection tolerance regions. It is seen that the CRM controllers adapt faster although without any excessive oscillations. They also enter the tolerance region but never exceed the projection boundary.

\clearpage

\section{Summary}
In this paper, we introduce a complete model of a flexible quadrotor with uncertain dynamics. In addition, we conduct a human-in-the-loop stability analysis of the overall closed loop control system, consisting of the flexible UAV model, operator model with reaction time delay and a closed loop reference model (CRM) adaptive controller, and provide a delay-dependent stability condition. We then demonstrate, via a simulation example, that the CRM adaptive controller not only can handle the uncertainties but also provides a smoother response compared to a conventional adaptive controller. Response characteristics are especially important for flexible systems due to the danger of excessive oscillations. Finally, we investigate the effect of the operator dynamics on the stability of the closed loop system, using the delay-dependent stability condition, for the specific simulation example.

% ***********************************************

\section*{Disclosure statement}
No potential conflict of interest was reported by the authors.

\section*{Funding}
This research was sponsored by the Scientific and Technological
Research Council of Turkey under Grant 118E937.

\appendix

\section{The Normalization Constant}
\label{app:AppB}
Let
\begin{equation}\label{betaapp}
	\bar{\beta}_j^*= \frac{\cos \bar{\beta}_j+\cosh \bar{\beta}_j}{\sin \bar{\beta}_j+\sinh \bar{\beta}_j}
\end{equation}
Substituting \eqref{betaapp} into \eqref{W3}, we obtain
\begin{equation}\label{W3upd}
	W_j(\bar{x})=\bar{\gamma}_j\left[\left(\cos \beta_j \bar{x}-\cosh \beta_j \bar{x}\right)-\bar{\beta}_j^* \left(\sin \beta_j \bar{x}-\sinh \beta_j \bar{x}\right),\right]	
\end{equation}
The normalization constant $\bar{\gamma}_j$ in \eqref{W3upd} can be calculated by the orthogonality of mode shape $W_j(\bar{x})$ as
\begin{equation}\label{normy}
	\int_{0}^{L_c} \rho_c A_c W_j^2(\bar{x}) d\bar{x} = 1,
\end{equation}
\begin{equation}\label{thisintegral}
	\int_{0}^{L_c} \rho_c A_c  \bar{\gamma}_j^2  [\left(\cos \beta_j \bar{x}-\cosh \beta_j \bar{x}\right)-\bar{\beta}_j^*\left(\sin \beta_j \bar{x}-\sinh \beta_j \bar{x}\right)]^2 d\bar{x} = 1.
\end{equation}
Solving \eqref{thisintegral}, the normalization constant is obtained as
\begin{equation}
	\bar{\gamma}_j=\frac{1}{\sqrt{\rho_c A_c \gamma_c}},
\end{equation}
where
\begin{equation}
	\begin{aligned}
		\begin{split}
			\gamma_c= \frac{1}{4\beta}[
			&-\bar{\beta}_j^{*2} \sin(2\bar{\beta}_j) +
			\bar{\beta}_j^{*2} \sinh(2\bar{\beta}_j) + 4 \bar{\beta}_j^{*2} \cos(\bar{\beta}_j) \sinh(\bar{\beta}_j)\\
			&-4 (\bar{\beta}_j^{*2} + 1) \sin(\bar{\beta}_j) \cosh(\bar{\beta}_j)
			+ 2 \bar{\beta}_j^{*} \cos(2\bar{\beta}_j)
			-2 \bar{\beta}_j^{*} \cosh(2\bar{\beta}_j)\\
			&+8 \bar{\beta}_j^{*} \sin(\bar{\beta}_j) \sinh(\bar{\beta}_j) + 4\bar{\beta}_j + \sin(2\bar{\beta}_j) + \sinh(2\bar{\beta}_j)
			-4 \cos(\bar{\beta}_j) \sinh(\bar{\beta}_j)].	
		\end{split}
	\end{aligned}
\end{equation}

\section{The Application of the Orthogonality Conditions}
\label{app:AppC}
Recall that the partial differential equations of motion for a damped Euler-Bernoulli beam is given as
\begin{equation}\label{pdeapp}
	E_c J_c \frac{\partial ^{4} w(\bar{x},t)}{\partial \bar{x}^{4}} + \rho_c A_c \frac{\partial ^{2} w(\bar{x},t)}{\partial t^{2}} + \sigma_c \frac{\partial  w(\bar{x},t)}{\partial t} = F(\bar{x},t).
\end{equation}
Using \eqref{gendisp} and applying separation of variables, it can be obtained that
\begin{equation}\label{sepapp}
	E_c J_c \frac{d^{4} W_j(\bar{x})}{d \bar{x}^{4}}= \rho_c A_c \bar{\omega}_j^2 W_j(\bar{x}). 
\end{equation}
Substituting \eqref{sepapp} into \eqref{pdeapp} and using \eqref{gendisp}, it follows that
\begin{equation}\label{einstein}
	\rho_c A_c \bar{\omega}_j^2 W_j(\bar{x}) \Upsilon_j(t) + 
	\rho_c A_c W_j(\bar{x}) \ddot{\Upsilon}_j(t)+
	\sigma_c W_j(\bar{x}) \dot{\Upsilon}_j(t)= F(\bar{x},t).
\end{equation}
Recall that the orthogonality conditions can be written as
\begin{equation}\label{orthoapp}
	\int_{0}^{L_c} \rho_c A_c W_j(\bar{x}) W_l(\bar{x}) d\bar{x} = \delta_{jl},
\end{equation}
where $\delta_{jl}$ is the Kronecker delta. Multiplying \eqref{einstein} by $W_l(\bar{x})$ and integrating it from 0 to $L_c$, it is obtained that
\begin{equation}\label{einstein2}
	\begin{aligned}
		\begin{split}
			\sum_{j=1}^{\infty}\bar{\omega}_j^2\Upsilon_j(t)\int_{0}^{L_c}\rho_c A_c  W_j(\bar{x}) W_l(\bar{x})d\bar{x}  &+ \\
			\sum_{j=1}^{\infty}		\ddot{\Upsilon}_j(t) \int_{0}^{L_c} \rho_c A_c W_j(\bar{x})W_l(\bar{x})d\bar{x} &+\\
			\frac{\sigma_c}{\rho_c A_c} \sum_{j=1}^{\infty}		\dot{\Upsilon}_j(t) \int_{0}^{L_c} \rho_c A_c W_j(\bar{x})W_l(\bar{x})d\bar{x} &=\int_{0}^{L_c} W_l(\bar{x})F(\bar{x},t)d\bar{x}. 
		\end{split}
	\end{aligned}
\end{equation}
In view of the orthogonality conditions given by \eqref{orthoapp}, it is obtained that
\begin{equation}\label{modaleqapp}
	\ddot{\Upsilon}_j(t)+\sigma_c'\dot{\Upsilon}_j(t)+\bar{\omega}_j^2\Upsilon_j(t)=\int_{0}^{L_c}W_j(\bar{x})F(\bar{x},t)d\bar{x},
\end{equation}
where $\sigma_c'=\sigma_c/(\rho_c A_c)$ is a constant.


\clearpage
\begin{thebibliography}{}

\bibitem[Alan, Yildiz, \& Poyraz(2018)]{alan2018}
Alan, A., Yildiz, Y., \& Poyraz, U. (2018). High-Performance Adaptive Pressure Control in the Presence of Time Delays: Pressure Control for Use in Variable-Thrust Rocket Development. IEEE Control Systems Magazine, 38(5), 26-52.	

\bibitem[Albaba, \& Yildiz(2019)]{albaba2019}
Albaba, B. M., \& Yildiz, Y. (2019). Modeling cyber-physical human systems via an interplay between reinforcement learning and game theory. Annual Reviews in Control, 48, 1-21.

\bibitem[Annaswamy, \& Gaudio(2020)]{anna2015}
Annaswamy A.M., Gaudio J.E. (2020) Robust Adaptive Control. In: Baillieul J., Samad T. (eds) Encyclopedia of Systems and Control. Springer, London.

\bibitem[Annaswamy, \& Yildiz(2020)]{anna2020enc}
Annaswamy A.M., Yildiz Y. (2020) Cyber-Physical-Human Systems. In: Baillieul J., Samad T. (eds) Encyclopedia of Systems and Control. Springer, London.

\bibitem[Baghdadi, Lowenberg, \& Isikveren(2011)]{bagh2011}
Baghdadi, N., Lowenberg, M. H., \& Isikveren, A. T. (2011). Analysis of flexible aircraft dynamics using bifurcation methods. Journal of Guidance, Control, and Dynamics, 34(3), 795-809.

\bibitem[Bauchau, \& Craig(2009)]{bau2009}
Bauchau, O. A., \& Craig, J. I. (2009). Structural analysis: with applications to aerospace structures (Vol. 163). Springer Science \& Business Media.

\bibitem[Bouabdallah(2007)]{bou2007}
Bouabdallah, S. (2007). Design and control of quadrotors with application to autonomous flying (No. THESIS). Epfl.

\bibitem[Cesnik et~al(2010)]{cesnik2010}
Cesnik, C., Senatore, P., Su, W., Atkins, E., Shearer, C., \& Pitchter, N. (2010). X-HALE: a very flexible UAV for nonlinear aeroelastic tests. In 51st AIAA/ASME/ASCE/AHS/ASC Structures, Structural Dynamics, and Materials Conference 18th AIAA/ASME/AHS Adaptive Structures Conference 12th (p. 2715).

\bibitem[Chang, Hodges, \& Patil(2008)]{chang2008}
Chang, C. S., Hodges, D. H., \& Patil, M. J. (2008). Flight dynamics of highly flexible aircraft. Journal of Aircraft, 45(2), 538-545.

\bibitem[Da Ronch, Badcock, Wang, Wynn, \& Palacios(2012)]{daronch2012}
Da Ronch, A., Badcock, K., Wang, Y., Wynn, A., \& Palacios, R. (2012, August). Nonlinear model reduction for flexible aircraft control design. In AIAA Atmospheric Flight Mechanics Conference (p. 4404).

\bibitem[Duhamel, \& Vetterli(1990)]{duhamel1990}
Duhamel, P., \& Vetterli, M. (1990). Fast Fourier transforms: a tutorial review and a state of the art. Signal Processing (Elsevier), 19(ARTICLE), 259-299.

\bibitem[Dussart, Portapas, Pontillo, \& Lone(2018)]{dussart2018}
Dussart, G., Portapas, V., Pontillo, A., \& Lone, M. (2018). Flight dynamic modelling and simulation of large flexible aircraft. Flight Physics-Models, Techniques and Technologies.

\bibitem[Dydek, Jain, Jang, Annaswamy, \& Lavretsky(2006)]{dydek2006}
Dydek, Z., Jain, H., Jang, J., Annaswamy, A., \& Lavretsky, E. (2006, August). Theoretically verifiable stability margins for an adaptive controller. In AIAA Guidance, Navigation, and Control Conference and Exhibit (p. 6416).

\bibitem[Dydek, Annaswamy, \& Lavretsky(2012)]{dydek2012}
Dydek, Z. T., Annaswamy, A. M., \& Lavretsky, E. (2012). Adaptive control of quadrotor UAVs: A design trade study with flight evaluations. IEEE Transactions on control systems technology, 21(4), 1400-1406.

\bibitem[Engelborghs, Luzyanina, \& Roose(2002)]{engelborghs2012}
Engelborghs, K., Luzyanina, T., \& Roose, D. (2002). Numerical bifurcation analysis of delay differential equations using DDE-BIFTOOL. ACM Transactions on Mathematical Software (TOMS), 28(1), 1-21.

\bibitem[Eraslan, Yildiz, \& Annaswamy(2020)]{eraslan2019}
Eraslan, E., Yildiz, Y., \& Annaswamy, A. M. (2020). Shared Control Between Pilots and Autopilots: Illustration of a Cyber-Physical Human System. IEEE Control Systems Magazine, 40(6), 79-99.

\bibitem[Feng, Wiltsche, Humphrey \& Topcu (2016)]{feng2016}
Feng, L., Wiltsche, C., Humphrey, L., \& Topcu, U. (2016). Synthesis of human-in-the-loop control protocols for autonomous systems. IEEE Transactions on Automation Science and Engineering, 13(2), 450-462.

\bibitem[Flatus(1992)]{flatus92}
Flatus, D. H. (1992). Aeroelastic stability of slender, spinning missiles. Journal of guidance, control, and dynamics, 15(1), 144-151.

\bibitem[Friswell, \& Lees(2001)]{fris2001}
Friswell, M. I., \& Lees, A. W. (2001). The modes of non-homogeneous damped beams. Journal of Sound and Vibration, 242(2), 355-361.

\bibitem[Gibson, Annaswamy, \& Lavretsky(2012)]{gibson2012}
Gibson, T., Annaswamy, A., \& Lavretsky, E. (2012). Improved transient response in adaptive control using projection algorithms and closed loop reference models. In AIAA Guidance, Navigation, and Control Conference (p. 4775).

\bibitem[Gibson, Annaswamy, \& Lavretsky(2013a)]{gibson2013a}
Gibson, T. E., Annaswamy, A. M., \& Lavretsky, E. (2013). On adaptive control with closed-loop reference models: transients, oscillations, and peaking. IEEE Access, 1, 703-717.

\bibitem[Gibson, Annaswamy, \& Lavretsky(2013b)]{gibson2013b}
Gibson, T. E., Annaswamy, A. M., \& Lavretsky, E. (2013). Closed-loop reference models for output-feedback adaptive systems. In 2013 European Control Conference (ECC) (pp. 365-370). IEEE.

\bibitem[Gibson(2014)]{gibsonphd2014}
Gibson, T. E. (2014). Closed-loop reference model adaptive control: with application to very flexible aircraft (Doctoral dissertation, Massachusetts Institute of Technology).	
	
\bibitem[Gürgöze, \& Erol(2006)]{gurgoze2006}
Gürgöze, M., \& Erol, H. (2006). Dynamic response of a viscously damped cantilever with a viscous end condition. Journal of Sound and Vibration, 298(1-2), 132-153.

\bibitem[Hesse, Palacios, \& Murua(2014)]{hesse2014}
Hesse, H., Palacios, R., \& Murua, J. (2014). Consistent structural linearization in flexible aircraft dynamics with large rigid-body motion. AIAA journal, 52(3), 528-538.

\bibitem[Islam, Liu, \& El Saddik(2017)]{islam2018}
Islam, S., Liu, P. X., \& El Saddik, A. (2017). Nonlinear robust adaptive sliding mode control design for miniature unmanned multirotor aerial vehicle. International Journal of Control, Automation and Systems, 15(4), 1661-1668.

\bibitem[ Khaitan, \& McCalley(2014)]{khai2014}
Khaitan, S. K., \& McCalley, J. D. (2014). Design techniques and applications of cyberphysical systems: A survey. IEEE Systems Journal, 9(2), 350-365.

\bibitem[Kreisselmeier, \& Anderson(1986)]{kreissel1986}
Kreisselmeier, G., \& Anderson, B. (1986). Robust model reference adaptive control. IEEE Transactions on Automatic Control, 31(2), 127-133.

\bibitem[L'Afflitto, \& Blackford(2020)]{laf2020}
L'Afflitto, A., \& Blackford, T. A. (2020). Constrained dynamical systems, robust model reference adaptive control, and unreliable reference signals. International Journal of Control, 93(5), 1039-1052.

\bibitem[Lavretsky(2011)]{lav2011}
Lavretsky, E. (2011). Reference dynamics modification in adaptive controllers for improved transient performance. In AIAA guidance, navigation, and control conference (p. 6200).

\bibitem[Lavretsky \& Wise(2013)]{lavretsky2013}
Lavretsky, E., \& Wise, K. A. (2013). Robust and adaptive control with aerospace applications.

\bibitem[Mahmoodi, Khadem, \& Kokabi(2007)]{mah2007}
Mahmoodi, S. N., Khadem, S. E., \& Kokabi, M. (2007). Non-linear free vibrations of Kelvin–Voigt visco-elastic beams. International Journal of Mechanical Sciences, 49(6), 722-732.

\bibitem[Meirovitch \& Nelson(1966)]{mei66}
Meirovitch, L., \& Nelson, H. D. (1966). On the high-spin motion of a satellite containing elastic parts. Journal of Spacecraft and Rockets, 3(11), 1597-1602.

\bibitem[Miller(2011)]{mill2011}
Miller, C. (2011, August). Nonlinear dynamic inversion baseline control law: architecture and performance predictions. In AIAA Guidance, Navigation, and Control Conference (p. 6467).

\bibitem[Narendra \& Valavani(1979)]{narendra1979}
Narendra, K. S., \& Valavani, L. S. (1979). Direct and indirect model reference adaptive control. Automatica, 15(6), 653-664.

\bibitem[Narendra \& Annaswamy(2012)]{narendra2012}
Narendra, K. S., \& Annaswamy, A. M. (2012). Stable adaptive systems. Courier Corporation.

\bibitem[Nguyen \& Tuzcu(2009)]{ngu2009}
Nguyen, N., \& Tuzcu, I. (2009, August). Flight dynamics of flexible aircraft with aeroelastic and inertial force interactions. In AIAA Atmospheric Flight Mechanics Conference (p. 6045).

\bibitem[Rao(2007)]{rao2007}
Rao, S. S. (2007). Vibration of continuous systems (Vol. 464). New York: Wiley.
	
\bibitem[Rasti \& Fazelzadeh(2012)]{rasti12}
Rasti, A., \& Fazelzadeh, S. A. (2012). Multibody dynamic modeling and flutter analysis of a flexible slender vehicle. International Journal of Structural Stability and Dynamics, 12(06), 1250049.

\bibitem[Ritter, Jones, \& Cesnik(2016)]{ritt2016}
Ritter, M., Jones, J., \& Cesnik, C. E. (2016). Enhanced Modal Approach for Free-Flight Nonlinear Aeroelastic Simulation of Very Flexible Aircraft. In 15th Dynamics Specialists Conference (p. 1794).

\bibitem[Romaszko, Sapiński, \& Sioma(2015)]{romas2015}
Romaszko, M., Sapiński, B., \& Sioma, A. (2015). Forced vibrations analysis of a cantilever beam using the vision method. Journal of Theoretical and Applied Mechanics, 53.

\bibitem[Sabatino(2015)]{sab2015}
Sabatino, F. (2015). Quadrotor control: modeling, nonlinearcontrol design, and simulation.

\bibitem[Schmidt(1998)]{schmidt98}
Schmidt, L. V. (1998). Introduction to aircraft flight dynamics. American Institute of Aeronautics and Astronautics.

\bibitem[Sowe, Simmon, Zettsu, de Vaulx, \& Bojanova(2016)]{sowe2016}
Sowe, S. K., Simmon, E., Zettsu, K., de Vaulx, F., \& Bojanova, I. (2016). Cyber-physical-human systems: Putting people in the loop. IT professional, 18(1), 10-13.

\bibitem[Srikanth, Annaswamy, \& Lavretsky(2010)]{srikant2010}
Srikanth, M., Annaswamy, A., \& Lavretsky, E. (2010, August). Dynamic modeling and control of a flexible four-rotor UAV. In AIAA Guidance, Navigation, and Control Conference (p. 8050).

\bibitem[Stengel(2015)]{steng2015}
Stengel, R. F. (2015). Flight dynamics. Princeton University Press.

\bibitem[Stepanyan, \& Krishnakumar(2010)]{step2010a}
Stepanyan, V., \& Krishnakumar, K. (2010). MRAC revisited: guaranteed performance with reference model modification. In Proceedings of the 2010 American Control Conference (pp. 93-98). IEEE.

\bibitem[Stepanyan, \& Krishnakumar(2011)]{step2011b}
Stepanyan, V., \& Krishnakumar, K. (2011). M-MRAC for nonlinear systems with bounded disturbances. In 2011 50th IEEE Conference on Decision and Control and European Control Conference (pp. 5419-5424). IEEE.

\bibitem[Thomsen, Annaswamy \& Lavretsky(2019)]{thomsen2019}
Thomsen, B. T., Annaswamy, A. M., \& Lavretsky, E. (2019). Shared control between adaptive autopilots and human operators for anomaly mitigation. IFAC-PapersOnLine, 51(34), 353-358.

\bibitem[Thurling(2000)]{thur2000}
Thurling, A. J. (2000). Improving UAV handling qualities using time delay compensation (No. AFIT/GAB/ENY/00M-01). Air Force Inst of Tech Wright-Patterson AFB OH.

\bibitem[Tohidi, Yildiz \& Kolmanovsky(2020)]{tohidi2020}
Tohidi, S. S., Yildiz, Y., \& Kolmanovsky, I. (2020). Adaptive control allocation for constrained systems. Automatica, Volume 121, 2020, 109161, ISSN 0005-1098.

\bibitem[Tran, Ge, \& He(2018)]{tran2018}
Tran, T. T., Ge, S. S., \& He, W. (2018). Adaptive control of a quadrotor aerial vehicle with input constraints and uncertain parameters. International Journal of Control, 91(5), 1140-1160.

\bibitem[Tullu, Byun, Kim \& Kang(2018)]{tullu2018}
Tullu, A., Byun, Y., Kim, J. N., \& Kang, B. S. (2018). Parameter optimization to avoid propeller-induced structural resonance of quadrotor type unmanned aerial vehicle. Composite Structures, 193, 63-72.

\bibitem[Van Schoor, \& von Flotow(1990)]{vanschoor90}
Van Schoor, M. C., \& von Flotow, A. H. (1990). Aeroelastic characteristics of a highly flexible aircraft. Journal of Aircraft, 27(10), 901-908.

\bibitem[Vepa(2014)]{vepa2014}
Vepa, R. (2014). Flight Dynamics, Simulation, and Control: For Rigid and Flexible Aircraft. CRC Press.

\bibitem[Verbeke \& Debruyne(2016)]{verbeke2016}
Verbeke, J., \& Debruyne, S. (2016). Vibration analysis of a UAV multirotor frame. In Proceedings of ISMA 2016 International Conference on Noise and Vibration Engineering (pp. 2401-2409).

\bibitem[Vinh(1995)]{vinh95}
Vinh, N. X. (1995). Flight mechanics of high-performance aircraft (Vol. 4). Cambridge University Press.

\bibitem[Waszak, Davidson, \& Schmidt(1987)]{wasznasa87}
Waszak, M. R., Davidson, J. B., \& Schmidt, D. K. (1987). A simulation study of the flight dynamics of elastic aircraft. Volume 1: Experiment, results and analysis.

\bibitem[Waszak \& Schmidt(1988)]{wasz88}
Waszak, M. R., \& Schmidt, D. K. (1988). Flight dynamics of aeroelastic vehicles. Journal of Aircraft, 25(6), 563-571.

\bibitem[Whitehead, \& Bieniawski(2010)]{whitehead2010}
Whitehead, B., \& Bieniawski, S. (2010, August). Model reference adaptive control of a quadrotor UAV. In AIAA Guidance, Navigation, and Control Conference (p. 8148).

\bibitem[Witte(2004)]{witt2004}
Witte, J. B. (2004). An investigation relating longitudinal pilot-induced oscillation tendency rating to describing function predictions for rate-limited actuators. Air Force Institute of Tech Wright-Patterson AFB OH School of Engineering and Management.

\bibitem[Wu \& Michiels(2012)]{wu2012}
Wu, Z., \& Michiels, W. (2012). Reliably computing all characteristic roots of delay differential equations in a given right half plane using a spectral method. Journal of Computational and Applied Mathematics, 236(9), 2499-2514.

\bibitem[Yildiz, Annaswamy, Yanakiev, \& Kolmanovsky(2010)]{yildiz2010}
Yildiz, Y., Annaswamy, A. M., Yanakiev, D., \& Kolmanovsky, I. (2010). Spark ignition engine fuel-to-air ratio control: An adaptive control approach. Control Engineering Practice, 18(12), 1369-1378.

\bibitem[Yucelen, De La Torre, \& Johnson(2014)]{yucelen2014}
Yucelen, T., De La Torre, G., \& Johnson, E. N. (2014). Improving transient performance of adaptive control architectures using frequency-limited system error dynamics. International Journal of Control, 87(11), 2383-2397.

\bibitem[Yucelen, Yildiz, Sipahi, Yousefi \& Nguyen(2017)]{yucelen2017}
Yucelen, T., Yildiz, Y., Sipahi, R., Yousefi, E., \& Nguyen, N. T. (2017). Stability analysis of human-adaptive controller interactions. In AIAA Guidance, Navigation, and Control Conference (p. 1493).


\end{thebibliography}
\end{document}